\documentclass[transmag]{IEEEtran}
\usepackage{latexsym}
\usepackage{graphicx}
\usepackage{amsfonts,amssymb,amsmath}
\usepackage{hyperref}
\usepackage[ruled,vlined]{algorithm2e}

\usepackage{relsize,xcolor}
\usepackage{bbm}
\usepackage{amsthm}
\usepackage{enumitem}
\usepackage{microtype}
\usepackage{subfigure}
\usepackage{booktabs} 
\usepackage{makecell}
\usepackage{xpatch}
\usepackage{breqn}
\usepackage{cite}
\usepackage{url}

\def\BibTeX{{\rm B\kern-.05em{\sc i\kern-.025em b}\kern-.08em T\kern-.1667em\lower.7ex\hbox{E}\kern-.125emX}}
\newcommand{\BB}{\mathbf{\mathcal{B}}}
\newcommand{\ints}{\mathlarger{\int}}
\newcommand{\tab}{\qquad\qquad\qquad}
\newcommand{\ra}{\rightarrow}
\newcommand{\norm}[1]{\left\lVert#1\right\rVert}

\newcommand{\ThmWDT}{1}
\newcommand{\ThmWQT}{2}
\newcommand{\ThmSWQT}{3}
\newcommand{\ThmKS}{4}
\newcommand{\ThmMMD}{5}
\newcommand{\ThmCM}{A.1}
\newcommand{\ThmSlutsky}{A.2}
\newcommand{\ThmPort}{A.3}
\newcommand{\ThmGC}{A.4}
\newcommand{\ThmRamdas}{A.5}
\mathchardef\mhyphen="2D
\newcommand{\algrule}[1][.6pt]{\par\vskip.5\baselineskip\hrule height #1\par\vskip.5\baselineskip}

\newtheorem*{theorem*}{Theorem}

\newtheorem*{corollary*}{Corollary}
\newtheorem*{definition*}{Definition}

\begin{document}

\title{On Matched Filtering for Statistical Change Point Detection}

\author{Kevin C. Cheng$^\dagger$, Eric L. Miller, Michael C. Hughes, Shuchin Aeron$^\dagger$
\thanks{
This research was funded by the NSF through TRIPODS grant 1934553 and U.S. Army Combat Capabilities Development Command Soldier Center (CCDC Soldier Center) via Tufts Center for Applied Brain and Cognitive Sciences (CABCS) under ARM994. Shuchin Aeron was also supported in part by NSF CCF:1553075 and AFOSR FA9550-18-1-0465. $^\dagger$ \textbf{Corresponding authors}: Kevin Cheng, Shuchin Aeron (shuchin@ece.tufts.edu)}}

\IEEEtitleabstractindextext{\begin{abstract}Non-parametric and distribution-free two-sample tests have been the foundation of many change point detection algorithms. However, randomness in the test statistic as a function of time makes them susceptible to false positives and localization ambiguity. We address these issues by deriving and applying filters matched to the \textit{expected} temporal signatures of a change for various sliding window, two-sample tests under IID assumptions on the data.  These filters are derived asymptotically with respect to the window size for the Wasserstein quantile test, the Wasserstein-1 distance test, Maximum Mean Discrepancy  squared (MMD$^2$), and the Kolmogorov-Smirnov (KS) test. 
The matched filters are shown to have two important properties.
First, they are distribution-free, and thus can be applied without prior knowledge of the underlying data distributions.
Second, they are peak-preserving, which allows the filtered signal produced by our methods to maintain expected statistical significance.
Through experiments on synthetic data as well as activity recognition benchmarks, we demonstrate the utility of this approach for mitigating false positives and improving the test precision.  Our method allows for the localization of change points without the use of \textit{ad-hoc} post-processing to remove redundant detections common to current methods. We further highlight the performance of statistical tests based on the Quantile-Quantile (Q-Q) function and show how the invariance property of the Q-Q function to order-preserving transformations allows these tests to detect change points of different scales with a single threshold within the same dataset.
\end{abstract}

}

\maketitle

\section{Introduction} \label{sec:intro}
Given a time-varying signal, the problem of change point detection (CPD) is to identify specific points in time where the signal exhibits a significant change either in its deterministic content or underlying stochastic distribution. Having been studied for nearly a century by researchers, CPD was originally motivated in problems of fault detection and quality control \cite{shewhart_economic_1931}. Since then, a wide range of CPD methods has been developed and applied across a diverse set of applications including finance \cite{adams_bayesian_2007}, human activity analysis \cite{liu_change-point_2013}, ECG and EEG processing \cite{qi_novel_2014, dizaji_change-point_2017}, speech \cite{li_m-statistic_2015}, sensor networks \cite{ting_he_nonparametric_2006, ciuonzo_dechade_2018}, and climate change \cite{reeves_review_2007}.

A common framework for CPD is to compute a univariate statistic measuring the similarity between two windows of data to either side of a purported change point. 
Methods reliant on this \emph{similarity-based} framework process an entire signal by iteratively ``sliding'' the pair of adjacent windows forward in time over the data, computing a test statistic typically derived from the Empirical Distribution Functions (EDFs), Quantile Functions (QFs), or Quantile-Quantile (Q-Q) function associated wiht the data in a given pair of windows. Then a hypothesis test can be applied to this statistic where the null hypothesis posits that no change point exists between the current two windows and thus the empirical distributions of the adjacent windows are drawn from the same distribution. If the null is true, the derived statistic should be small in some sense. At a change point, the data in the two windows come from different distributions and thus the expected value of the derived statistic peaks. Therefore, it is common under this framework to only consider local maxima as candidate change points for hypothesis testing. If the distribution of the statistic's value under the null hypothesis is known, the null can be rejected and a change point declared with a certain confidence should the peak value exceed some corresponding threshold.  In contexts where the distribution under the null cannot be attained, this peak thresholding method can still be used to make detections but lacks formal guarantees. 

Two concepts central to the ideas in this paper are the notion of a \textit{distribution-free} test and a \textit{peak-preserving} transformation.  Non-parametric tests that are distribution-free have their probabilistic distribution of the test statistic under the null hypothesis independent of the distribution generating the data. Therefore, if a test is distribution-free, threshold values that correspond to rejection of the null at a fixed false-alarm rate can be applied regardless of the change in distribution to be detected; thus, one does not require methods such as density estimation to derive the distribution under the null.  Peak-preservation applies to any method that preserves the expected value of the test statistic at a candidate change point. Since the change point statistic is expected to peak at a true change point, any peak-preserving post-processing method will maintain the statistical significance of the test statistic at the change points.

One issue that is not well studied in multiple change-point detection is the exact determination of change points once the appropriate statistic is computed. Because the test statistic time series is itself random (since it is a function of the underlying random observations), one tends to see multiple local maxima in the vicinity of a change resulting in either a large number of false alarms or the need for \textit{ad-hoc} post processing to identify true change points. Current state-of-the-art methods simply consider local maxima above a specified threshold as change points~\cite{truong_selective_2020}, \cite{li_m-statistic_2015}, and remove duplicate detections within a specified window \cite{sugiyama_direct_2008}. However, it is clear that the sliding window methods produce a correlated test statistic where the effects of the change at a given point in time are spread over an interval which contains the change point. 

Motivated by this fact, we draw on the classical signal processing idea of a matched filter \cite{oppenheim_signals_2016} as a tool to better identify change points.
More specifically, the asymptotic (as window length goes to infinity) forms of the expected ``signatures” produced by sliding window methods are derived for the Wasserstein-1 Distance (W1-DT), Wasserstein Quantile Test (WQT), Sliced Wasserstein Quantile Test (SWQT), Maximum Mean Discrepancy squared (MMD$^2$), and Kolmogorov-Smirnov (KS) distance.  Under a commonly-used assumption that the data in each segment is independent and identically distributed (IID) \cite{adams_bayesian_2007, truong_selective_2020}, we propose a novel model for analyzing similarity-based tests in the sliding window framework using mixture models, and prove that the expected signature of each statistical test converges to a function that is independent (up to a scale factor) of the data distribution prior to and following the change point. Thus, distinct from the tests themselves being distribution-free, these filters are shown to be distribution-free in that they can be applied without knowing the distribution of the data. 

Using the asymptotically derived signature, we construct finite length filters in a manner that is peak-preserving. In summary of our main findings, the filters for the KS and W1-DT are piecewise linear while those of the WQT and MMD$^2$, which are based on a square distance, are quadratic. While the matched filters for these statistics are of linear or quadratic form, there is no reason to believe that matched filter derived for other statistics are guaranteed to be in these forms.

Matched filters are generally known to be time-reversed versions of the signal to be detected, and are considered optimal with respect to signal to noise ratio when, for example, the noise is additive, white, and Gaussian. We do not claim that the stochasticity in the test statistic for these sliding window methods is of the form for which the matched filtering process is in any sense optimal. In addition, the IID condition in which the matched filters are derived rarely applies to real-world data. In our evaluation, we first consider simulated data where samples are generated IID from scalar and multivariate Gaussians and show that the matched filter simplifies the peak-detection process and improves CPD performance when considering precision-recall based metrics. We then demonstrate similar performance improvements when extended to real-world activity data where the IID condition does not hold.

Finally, exploration of the performance of these filters across the five non-parametric test statistics identified above brings to light interesting properties of tests derived from the Quantile-Quantile (Q-Q) function, which include the WQT and SWQT.  Specifically, it is well known that the Q-Q function is both invariant to order-preserving transformations of the data \cite{wasserman_all_2006} and we show that it is also is highly sensitive to small changes in the support of the data. In the context of CPD, these properties allow Q-Q tests to detect relatively small changes in a manner that is practically independent of the overall scale of the data.  Whether this characteristic is useful or a source of false alarms depends heavily on the underlying application, an issue that is examined in this work using both simulated and real-world data. 

In summary, the main contributions of this paper are as follows: 
\begin{itemize}
    \item We develop a novel \emph{principled methodology} for deriving and applying matched filters for similarity-based change point detection .
    We prove that our proposed matched filters are \emph{distribution-free} and \emph{peak-preserving}, which preserves the guarantees provided by standard hypothesis tests when used to detect change points given the filtered signal.

    \item We offer \emph{formal proofs} deriving the asymptotically matched filter for four common statistical tests: the Kolmogorov-Smirnov test (KS), Maximum Mean Discrepancy squared (MMD$^2$), Wasserstein-1 distance (W1-DT), Wasserstein quantile test (WQT), and propose the sliced Wasserstein quantile test (SWQT) as a multivariate extension of the WQT. 

    \item We demonstrate \emph{empirical benefits} for the above theoretical contribution in both simulation studies and real-world multivariate change point benchmarks. We show how matched filters with suitable finite-length approximations can deliver improved performance evaluated using precision-recall based metrics including the reduction in false positives, and importantly, remove any need for additional post-processing for duplicate detection removal common to other methods.

    \item We provide \emph{insight} into how the choice of test statistic impacts empirical performance, specifically highlighting differences in sensitivity between statistics based on the Quantile-Quantile (Q-Q) function compared statistics derived from the Empirical Distribution Function (EDF) or Quantile Function (QF). These insights are justified through theory and demonstrated in simulated and real-world human activity datasets.
\end{itemize}

The remainder of this paper is organized as follows. 
Sec.~\ref{sec:problem} introduces the CPD problem and setup the basic framework for CPD with statistical tests on sliding windows. In addition, we outline the main properties that differentiate Q-Q tests from other statistical tests. 
Sec.~\ref{sec:MF} motivates and outlines our simple but novel approach of deriving and applying matched filters and states the main theorems for the asymptotically matched filters for WQT, MMD$^2$, KS, and W1-DT tests. Details of each proof are left to the appendix.
Sec.~\ref{sec:experiments} shows that the empirically computed matched filters and properties of Q-Q based tests match our theoretical results. Furthermore we demonstrate the improvement in detection as shown through false alarm rates, and the related metrics of precision and recall evaluating on simulated data as well as real-world benchmarks based on human activity (HASC \cite{ichino_hasc-pac2016:_2016}, MASTRE \cite{hussey_monitoring_2020}) and honeybee activity (Beedance) \cite{oh_learning_2008}.

\begin{figure*}
    \centering
    \includegraphics[width=1.8\columnwidth]{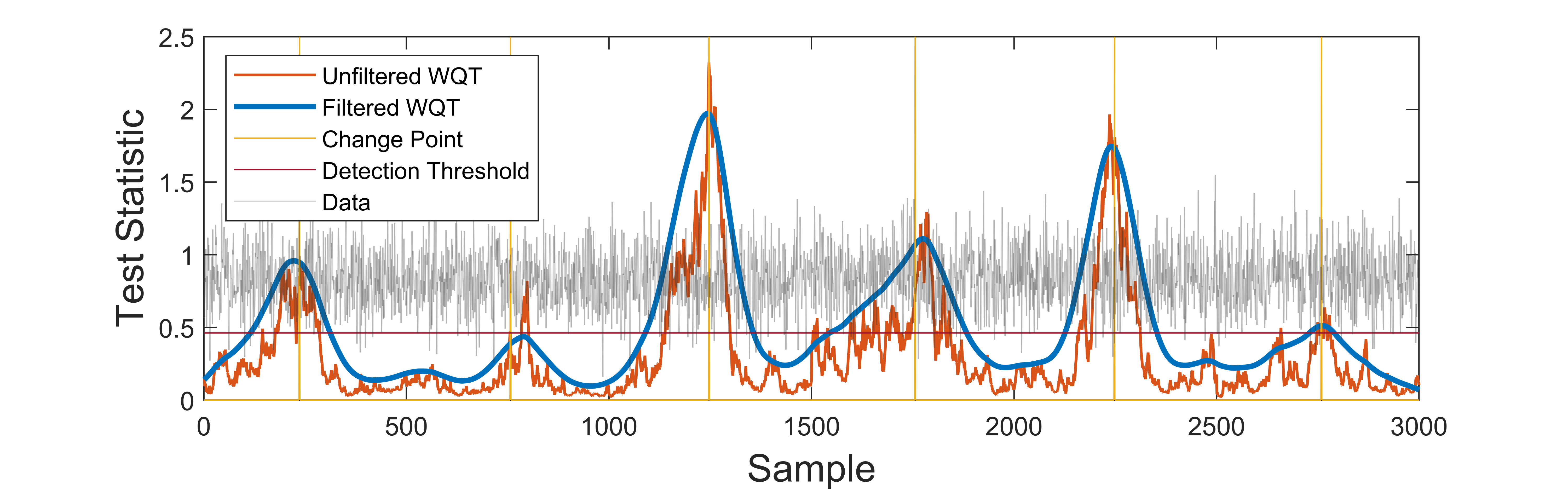}
    \caption{Result of change point detection using the WQT on sliding windows of size $n=150$ on simulated data outlined in \ref{sec:experiments}. The noisy unfiltered change point statistic can cause false detections, and complicate localization of change points. The data alternates between two normal distributions $\mathcal{N}(0,1), \mathcal{N}(0.25,1)$. The detection threshold corresponds to rejection of the null with 95\% confidence.}
    \label{fig:motivation}
\end{figure*}

\section{Related Work}

Change point detection methods are classified along a number of different dimensions \cite{truong_selective_2020, aminikhanghahi_survey_2017}, such as whether labels are available at training (supervised or unsupervised), the detection setting (offline or online), the number of change points assumed, the dimensionality of the signal, and the existence of modeling assumptions on the data (model-based or similarity-based).
\textit{Supervised} CPD methods require an annotated training corpus where each time-varying signal has associated, labeled change points. Once the method is ``trained'' on these examples, it can be used to process new signals whose change points are unknown. In contrast, \emph{unsupervised} CPD methods require only a set of signals to make detections.
\emph{Online} CPD methods make detections using only historical data and thus are useful in streaming or real-time settings. \emph{Offline} methods process an entire sequence retrospectively. 
\textit{Single change point} problems terminate after one change is detected while \textit{multiple change point} methods focus on time series where many changes may occur and are further divided based on whether or not the number of change points is known \textit{a-priori}.
While recent benchmarks~\cite{burg_evaluation_2020} indicate that many competitive CPD methods are only capable of processing univariate signals, a growing number can handle multivariate signals.

CPD methods can also be classified as either primarily model-based or similarity-based. 
We define \emph{model-based} techniques as those that make specific probabilistic assumptions about underlying distributions of data, such as by characterizing changes in the values of known \cite{chamroukhi_joint_2013} or learned \cite{lee_time_2018} parameters. Model-based efforts originate from the landmark works by \cite{wald_sequential_1947, page_continuous_1954, shewhart_economic_1931}.  The pioneering use of state-space dynamical models \cite{willsky_generalized_1976} formed the basis for work in hierarchical Bayesian models such as switching linear-dynamical systems (SLDS) \cite{murphy_switching_1998}. A competitive model-based method in recent benchmarks \cite{burg_evaluation_2020} is ``Bayesian Online Change Point Detection'' (BOCPD)~\cite{adams_bayesian_2007}, which estimates the probability of change at each time via a switching state model that assumes within a segment that data arises IID from an exponential family likelihood.
Other model-based efforts have pursued Bayesian \emph{nonparametric} approaches using Gaussian processes~\cite{saatci_gaussian_2010} to avoid this restrictive IID assumption and achieve more flexible within-segment data models.

Generally, model-based methods for CPD are effective when the modelling assumptions hold and are able to capture the key characteristics of the signal change. 
In contrast, \emph{similarity-based} methods employ a test statistic derived directly from computing some suitable similarity or ``distance'' between windows of data samples without underlying assumptions concerning the specific generating distribution of the data. Therefore, similarity-based methods can be applied even if the proper assumptions about transitions between segments or the data distribution within a single segment are unknown or not easily expressed in a tractable probabilistic model. 

For similarity-based methods, three broad classes of tests have been applied for CPD.  The first class is likelihood ratio tests of the estimated distribution functions from the respective windows \cite{sugiyama_direct_2008, liu_change-point_2013, ciuonzo_dechade_2018}. The second class is non-parametric statistical tests, such as the Kolmogorov-Smirnov (KS), Cramer-von Mises, and Mann-Whitney statistics \cite{hawkins_nonparametric_2010, ross_two_2012}. The last class is tests based on distance metrics between empirical probability distributions. Prominent among them is the family of integral probability metrics \cite{sriperumbudur_empirical_2012} which includes the Maximum Mean Discrepancy (MMD) \cite{gretton_kernel_2012}, and the Wasserstein distance~\cite{sommerfeld_inference_2016}, which is closely related to the Q-Q based Wasserstein Quantile Test (WQT) ~\cite{ramdas_wasserstein_2015, cheng_optimal_2020}.

Our proposed method expands on the similarity-based framework by applying matched filters to commonly used statistical tests, and is capable of handling multivariate signals for unsupervised, multiple CPD where the number of changes is not assumed known. This setting is motivated by intended applications in human activity analysis in the modern era, where large volumes of highly sampled sensor data are affordable but annotation is prohibitively expensive and suitable probabilistic assumptions for this data are challenging. To address CPD in these contexts we propose a simple but effective method that has few hyperparameters and makes minimal assumptions on the data.

\section{Change Point Detection}
\label{sec:cpd}\label{sec:problem}
\subsection{Problem Statement} 
Assume a time series, $X[t] \in \mathcal{C}, t = 1,2,...,$ where $\mathcal{C} \subset  \mathbb{R}^d$ represents a compact set, is constructed with the following model.
\begin{enumerate}
    \item The data consists of distinct time segments $[0, \tau_1], [\tau_1+1, \tau_2],...,[\tau_k+1, \tau_{k+1}],...$ with $\tau_1 < \tau_2 < ...$, such that within each time segment, $X[t], t \in [\tau_k+1, \tau_{k+1}]$ are IID samples from a fixed but \emph{unknown} distribution. 
    \item The distributions in successive time segments are different but in general two non-adjacent segments can have samples from the same distribution. 
\end{enumerate}
The set of time points $\{\tau_1, \tau_2, ... \}$, are referred to as the change points. Given these conditions, the problem of unsupervised \textbf{Change Point Detection (CPD)} is to estimate the (possibly empty) set of change points purely from the provided $X[t]$ data without any information or assumptions about the number or location of change points. 

\subsection{Notation and Background}
A set of $n$ total samples, $\mathcal{P}_n = \{X_1, X_2, ..., X_n\}$, each drawn IID from some distribution $P$, has an associated Empirical Distribution Function (EDF), and Quantile Function (QF) defined as,

\addtolength{\tabcolsep}{-5pt} 
\begin{align}\label{eq:EDF_QF_QQ_1}
    \centering
    \begin{tabular}{lcl}
EDF:&$P_n(x) \triangleq \frac{1}{n}\sum\limits_{i=1}^n \mathbbm{1}_{X_i \leq x}$  & $\mathcal{C} \ra [0,1]$ \\
QF: \hspace{5mm} &$P_n^{-1}(x) \triangleq \inf \{y: P_n(y)\geq x\}$ \hspace{5mm} &  $[0,1]\ra \mathcal{C}$ 
    \end{tabular}
\end{align}
\addtolength{\tabcolsep}{5pt} 
\noindent where the indicator function is,
\begin{align}
     \mathbbm{1}_{X_i \leq x} \triangleq \begin{cases} 1 & X_i \leq x \\ 0 & X_i > x.\end{cases}
\end{align} 
Given another set $\mathcal{Q}_m$ of $m$ IID samples, drawn from distribution $Q$, the Quantile-Quantile (Q-Q) function is as follows,
\begin{align}\label{eq:EDF_QF_QQ_2}
    \centering
    \begin{tabular}{cl}
$P_n(Q_m^{-1})(x) \triangleq\frac{1}{n}\sum\limits_{i=1}^n \mathbbm{1}_{X_i \leq \inf \{y: Q_m(y)\geq x\} }$ & \hspace{2mm} $[0,1] \ra [0,1]$.
    \end{tabular}
\end{align}
The EDF, QF and Q-Q functions represent stochastic processes on their respective domains \cite{billingsley_convergence_1999}. Thus in this work, almost sure convergence ($\ra_{as}$) and weak convergence or convergence in distribution ($\ra_w$), refers to convergence of stochastic processes. While the relevant background is covered in the appendix, we point the readers to additional works \cite{vaart_asymptotic_1998, van_der_vaart_weak_1996, billingsley_convergence_1999, pollard_convergence_2012} for more comprehensive coverage in this area. 

\subsection{Quantile-Quantile Tests}
On real valued data, many of the two-sample tests discussed in this paper directly compare EDFs or QFs (Tab.~\ref{tab:TheTable}).  In the Q-Q case, the comparison is made to the uniform distribution. Indeed, when two distributions are equal, as under the CPD null, the resulting Q-Q function matches the uniform distribution on $[0,1]$. Thus, we designate statistics of this form as \textit{Q-Q tests}.  

Since the Q-Q function is non-decreasing from $[0,1]\ra[0,1]$, Q-Q tests are inherently bounded.  The maximum is achieved when the open intervals covering the respective supports of $P$ and $Q$ are disjoint, as shown in Appx.~\ref{sec:WQTMax} for the WQT. Q-Q functions are also known to be invariant to any transformation on the data that is order-preserving, or in other words, monotonically increasing \cite{wasserman_all_2006}, such as positive affine transforms.

The bounded property of the Q-Q function makes Q-Q based tests particularly sensitive to small shifts in support. In addition, the invariance of Q-Q functions to order-preserving transformations makes Q-Q tests rather insensitive to such transformations of the data. More precisely, the Q-Q test statistic will be identical for a given a time series $X[t]$ and $f(X[t])$ where $f(x)$ is an order-preserving function.  For Q-Q tests applied to multiple change point detection problems, these properties imply that a single threshold can be used to detect changes of widely varying magnitudes; at least those changes induced by shifts in support or order-preserving transformations.  We verify this claim using simulated data in Section \ref{sec:SimulatedResults}.  In the case of real-world data, we discuss how the empirical results in Section \ref{sec:RealWorldData} may be interpreted in light of the properties of Q-Q test.

\begin{figure*}
    \centering
    \includegraphics[width=1.6\columnwidth]{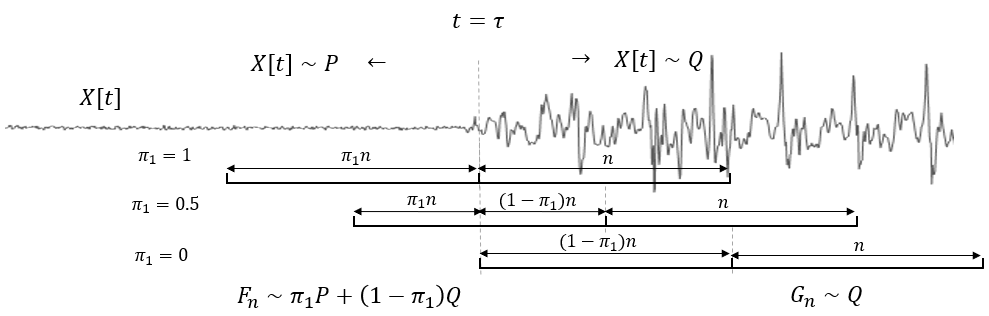}
    \caption{Diagram of the assumed modeling setting for the test statistic response as function of time around a true change point.
    \emph{Top:} The observed signal is assumed to arise from two distinct distributions, $P$ and $Q$, with change point at time $t = \tau$. Before $\tau$, data is sampled IID from $P$. After $\tau$, data is IID from $Q$.
    \emph{Bottom:} Our sliding window framework computes a test statistic between adjacent windows $F$ and $G$ of size $n$ at each time $t$. We consider the expected value of the statistic changes from the true change point $\tau$ moving right to $\tau + n$ (without loss of generality, moving left is the same as moving right assuming the proposed test is symmetric).
    The left window $F$ represents a \emph{mixture} of samples from $P$ and $Q$ with mixture proportion $\pi_1$.
    The right window $G$ will be purely sampled from $Q$.
    \emph{Takeaway:} Formally characterizing the test statistic response as a function of $\pi_1$ is the goal of our asymptotically matched filter.
    }
    \label{fig:diagram}
\end{figure*}

\subsection{Statistical Tests for Change Point Detection}\label{sec:statCPD}
Our general framework for CPD generates a test statistic by sliding adjacent windows of a constant size $n$ and computing a two-sample test from samples within each window (Fig.~\ref{fig:diagram}). 

At each time $t \geq n$, we define two windows of samples of size $n$; one to the left of $t$, $\{X[t-n], X[t - n +1],..,X[t-1]\}$ with distribution $F_n[t]$ and the other from the $n$ samples to the right $\{ X[t], X[t +1],..,X[t+(n-1)]\}$, with distribution $G_n[t]$. A statistical test $D_*(F_n[t], G_n[t])\,$ is then applied to these two windows. 
We use notation $D_*$ to represent a general statistical test. This can be substituted for the various specific statistical tests defined in Sec.~\ref{sec:MFAll}.

The nominal approach for identifying change points given a test statistic is to label local maxima of a computed statistic above some threshold parameter \cite{li_m-statistic_2015}. However, as is evident in Fig.~\ref{fig:motivation}, randomness in the statistic can adversely affect CPD methods that use thresholds for detection. Furthermore, in the presence of change points, multiple peaks complicate the exact localization of change points. These problems, along with the fact that sliding windows produce a correlated statistic, motivate the application of matched filtering of the test statistic for CPD.

\section{Matched Filters for Statistical Tests}\label{sec:MFAll}

\subsection{Asymptotically Matched Filters}\label{sec:MF}
As shown in Fig.~\ref{fig:diagram}, for the sliding window framework with constant window size $n$, the effect of a change point located at $t=\tau$ will be reflected in the test statistic on the interval $[\tau-n, \tau+(n-1)]$. Consequently, the matched filter is derived from the expected response of the test statistic on this interval. 

In this change point scenario, samples are assumed to be drawn IID from a distribution $P$ for $(\tau-2n) \leq t<\tau$, and from another distribution $Q$ for $\tau \leq t < (\tau+2n)$. We then generate the EDFs from adjacent sliding windows and the test statistic $D_*$ as described in section \ref{sec:statCPD}. Data in the windows that span the change point can be modeled as coming from a mixture distribution between $P$ and $Q$.

Without loss of generality (as long as $D_*$ is shown to be symmetric), we consider the case starting at $t=\tau$ and sliding the window to the \textit{right} such that the change point is located in the \textit{left} set of samples. In this setup, samples from the left window can be modeled as IID samples from the mixture $F_n[t] \sim \left( (1-\frac{t-\tau}{n}) P + \left(\frac{t-\tau}{n}\right)Q\right)$ and the distribution of the samples in the right window remains constant, $G_n[t] \sim Q$. We redefine these distributions in terms of a mixture parameter $\pi_1 = \left(1-\frac{(t-\tau)}{n}\right)$ where $t \in \{\tau, \tau+1, ..., \tau+n\}$ corresponds to $\pi_1 \in \{1, (1-\frac{1}{n}), ..., \frac{1}{n}, 0\}$.
In this mixture view, our left window $F_n$ and right window $G_n$ distributions are,
\begin{align}\label{eq:distr}
F_n(\pi_1) \sim \left(\pi_1 P + (1-\pi_1)Q\right), \hspace{5mm} G_n(\pi_1) \sim Q.
\end{align}
This reformulation allows us to analyze the expected response of the statistic asymptotically, as $n$ goes to infinity. In this case, the possible values of $\pi$ become countably infinite on the interval $[1,0]$.

Given that $F_n$ and $G_n$ are derived from $n$ samples drawn IID from the distributions defined in \eqref{eq:distr}, we prove that as $n \ra \infty$ for $\pi_1 \in [1,0]$, the expected value of each similarity statistic discussed in this paper converges to a deterministic function of the following form,
\begin{align} \label{eq:asymp}
\mathbb{E}[D_*( F_n(\pi_1), G_n(\pi_1))] \ra d_*(P,Q)  h(\pi_1).
\end{align}

\noindent Here, $d_*(P,Q)$ is a constant for a given $D_*(\cdot, \cdot), P$, and $Q$, while $h(\pi_1)$ is only a function of $\pi_1$. From \eqref{eq:asymp} we can conclude two main properties. First, the time dependent component of the response; that is $h$, is independent of the distributions $P$ and $Q$. Second, $P$ and $Q$ impact the response only through a constant scale factor. These two properties allow us to construct a matched filter whose shape is \textit{distribution-free} and \textit{peak-preserving} which we precisely define in Sec.~\ref{sec:PP_DF_MF}.

Moving back to the case where $n$ is finite, \eqref{eq:asymp} suggests that the expected value of the test statistic around a change point at $t=\tau$ for $t = (\tau-n), \dots, (\tau+n)$ be approximated as,
\begin{align}\label{eq:toOp}
    E\left[D_*(F_n[t], G_n[t])\right] \approx d_*(P,Q) h\left(1-\frac{|t-\tau|}{n}\right).
\end{align}
\noindent Since $D_*(\cdot, \cdot)$ is symmetric in its arguments, \eqref{eq:asymp} holds when $F_n(\pi_1), G_n(\pi_1)$ are reversed, which corresponds to the analogous setup with the windows slid to the left. Therefore, the response of the statistic is mirrored about $t=\tau$. For shorthand, change point statistic is denoted as $D_*[t] \triangleq D_*( F_n[t], G_n[t])$.

Therefore, with a slight abuse of notation, we can define the matched filter $h[t]$,
\begin{align}
    h[t] = \begin{cases} h\left(\pi_1 = \left(1-\frac{|t|}{n}\right)\right) &  -n \leq t \leq n \\
    0 & \text{otherwise.}
    \end{cases}
\end{align}

\label{sec:method_overview}
\label{sec:OT}
\begin{table*}[t]
    \centering
    \begin{tabular}{c|c|c|c|c|c }
    Test                    & WQT  & SWQT & W1-DT & KS & MMD$^2$ \\
    \hline
    \hline
    Type of Test & Q-Q & Q-Q  & QF, EDF & EDF & N/A \\
    \hline
    \makecell{Distribution \\Under Null} &  DF~\cite{ramdas_wasserstein_2015} & DF &  not DF~\cite{sommerfeld_inference_2016} & DF~\cite{vaart_asymptotic_1998} & not DF~\cite{gretton_kernel_2012} \\
    \hline
    Data Dimension & $\mathbb{R}^1$ & $\mathbb{R}^d$ & $\mathbb{R}^1$ & $\mathbb{R}^1$ & $\mathbb{R}^d$ \\
    \hline
    \makecell{\textbf{Matched filter} \\ $h[t]$, $t\in[-n, n]$ \\ (our contribution)} & $\left(1-\frac{|t|}{n}\right)^2$ & $\left(1-\frac{|t|}{n}\right)^2$ & $1-\frac{|t|}{n}$ & $1-\frac{|t|}{n}$ & $\left(1-\frac{|t|}{n}\right)^2$ 
    \end{tabular}
    \caption{\textnormal{Summary table comparing the two-sample tests discussed in this paper. EDF: empirical distribution function, QF: quantile function, Q-Q: quantile-quantile. DF: distribution-free. }}
    \label{tab:TheTable}
\end{table*}

\subsection{Peak-Preserving, Distribution-Free Matched Filters}\label{sec:PP_DF_MF}
The matched filter defined above can be applied to the test statistic signal $D_*[t]$ to produce a peak-preserving filtered signal $F_*[t]$ suitable for CPD:
\begin{align} \label{eq:optFilt}
    F_*[t]= D_*[t] \circledast \alpha h[t], \quad \alpha = \frac{1}{\sum_{t=-n}^n h[t]^2}. 
\end{align}
Here, $\alpha \geq 0$ is a scalar factor ensuring that the filter is in fact peak-preserving and $\circledast$ denotes the convolution operation.\footnote{boundary conditions handled through zero-padding} We note that this filtering process is distribution-free since \eqref{eq:optFilt} does not depend on $P$ or $Q$. This means that $h[t]$ is the matched filter regardless of the distributions defining the signal change and can be applied globally without any knowledge of probabilistic model of the change point.

It also follows from \eqref{eq:optFilt} that since the expected value of the test statistic $\mathbb{E}[D_*[t]]$ in the region of a change point at $t=\tau$ reflects \eqref{eq:asymp}, then $\mathbb{E}[D_*[t=\tau]] = \mathbb{E}[F_*[t=\tau]] = d_*(P,Q)$. Thus the resulting peak value of the statistic at the change point is preserved in expectation through the filtering process. This peak-preservation property is important for statistical tests where the resulting values are compared to a threshold in order to reject of the null hypothesis with a certain statistical confidence.

With the application of the matched filter, change points are detected at local maxima above a threshold of the filtered statistic, where no further post-processing of the local maxima is required. The proposed algorithm is detailed in full in Alg.~\ref{alg:algorithm}.

In the next few sections, we state the main theorem regarding the asymptotic expected value around a change point for each of the statistical tests of concern in this paper.  Proofs of these theorems are provided in Appxs.~\ref{app:W1-DT}--\ref{app:MMD}.

\begin{algorithm}\label{alg:algorithm}
\SetAlgoLined
 \SetKwInOut{Input}{Input}
 \SetKwInOut{Output}{Output}
 \Input{$X[t]:$ data, $t\in 1..T$ \\ $n:$ window size \\ $\eta:$ detection threshold \\ $D_*(\cdot, \cdot):$ statistical test, with \\ $h[t]:$ corresponding matched filter}
 \Output{$F_*[t]$: change point statistic \\ $\{\tau_1, \tau_2, \ldots \}$: change points}
 \For{$t=n:(T-n)$}{
   $F_n[t] \leftarrow \{X[t-n], ..., X[t-1]\}$\\
   $G_n[t] \leftarrow \{X[t], ..., X[t+(n-1)]\}$ \\
   $D_*[t] \leftarrow D_*(F_n[t], G_n[t])$
 }
 $\alpha \leftarrow \frac{1}{\left(\sum_{t=-n:n} h[t]^2\right)}$ \\
 $F_*[t] \leftarrow D_*[t] \circledast \alpha h[t]$ \\
 $\{\tau_1, \tau_2, \ldots \} \leftarrow \{t: F_*[t] > \max(\eta, F_*[t-1], F_*[t+1])\}$\\
\Indm\algrule
 \caption{Matched Filtered Statistical CPD}
\includegraphics[width=1.0\columnwidth]{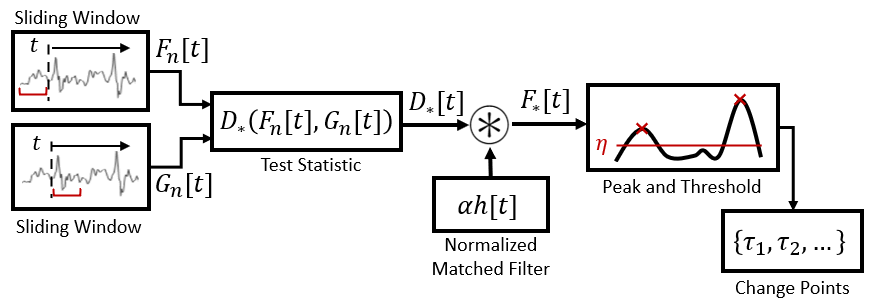}
\end{algorithm}

\subsection{Wasserstein-1 Distance Test (W1-DT)}\label{sec:WDT}
Given two probability distributions $F, G$ on $\mathbb{R}^d$ the Wasserstein-p distance $\mathcal{W}_p(F,G)$ is defined as, 
\begin{align} \label{eq:emd}
\mathcal{W}_p(F,G) = \left( \min_{\pi \in \Pi(F,G)} \mathbb{E}_{\pi(X,Y)} \left[ \norm{X-Y})^p \right] \right)^{\frac{1}{p}},
\end{align} 

\noindent where $\Pi(F,G)$ denotes the set of all joint distributions with marginals $F, G$. For $d=1, p=1$, the Wasserstein-1 distance has the closed form \cite{santambrogio_optimal_2015, peyre_computational_2018},
\begin{align}
    D_{w1\mhyphen dt}(F_n, G_n) \triangleq&  \int_0^1 |F_n^{-1}(x) - G_n^{-1}(x)| dx \nonumber \\
    =& \int_{\mathcal{C}} |F_n(x) - G_n(x)| dx.
\end{align}
We note the following theorem. 
\begin{theorem*} {\textnormal{\textbf{\ThmWDT:}}} \label{thm:wdt} Let $F_n$, $G_n$ be derived from IID samples drawn from $\left(\pi_1 P + (1-\pi_1) Q \right)$, and $Q$ respectively, where $P, Q$ are continuous on a compact domain, and constant $d_{w1\mhyphen dt}(P,Q) = \int_{\mathcal{C}} |(P(x)-Q(x))| dx$, then
\begin{align}
    D_{w1\mhyphen dt}(F_n, G_n)\ra_w \pi_1 d_{w1\mhyphen dt}(P,Q).
\end{align}
\end{theorem*}

Assuming that the samples live on a compact set, the test statistic is bounded. Thus it follows from the Portmanteau theorem \cite{billingsley_convergence_1999} that $\mathbb{E}\big[D_{wdt\mhyphen 1}(F_n, G_n)\big] \ra \pi_1 d_{wdt\mhyphen 1}(P,Q)$, which has the form of \eqref{eq:asymp}. Thus by the process described in Sec.~\ref{sec:MF} the matched filter for the operational case when $n$ is sufficiently large (but finite) is a piecewise linear function of $t$, $h[t]= \left(1-\frac{|t|}{n}\right)$.

\subsection{Wasserstein Quantile Test (WQT)} \label{sec:WQT}
The WQT  is a distribution-free variant of the Wasserstein distance that measures the Wasserstein distance of the Q-Q function to the uniform measure \cite{ramdas_wasserstein_2015},
\begin{align} \label{eq:WQTA}
    D_{wqt}(F_n,G_n) \triangleq & \frac{n}{2} \mathcal{W}_2(F_n G_n^{-1}, U[0,1])^2 \nonumber \\
    =& \frac{n}{2} \int_0^1 (F_n(G_n^{-1}(x))-x)^2 dx.
\end{align}

For the WQT we note  the following theorem. 
\begin{theorem*} {\textnormal{\textbf{\ThmWQT:}}} \label{thm:wqt}
Let $F_n$, $G_n$ be derived from IID samples drawn from $\left(\pi_1 P + (1-\pi_1) Q \right)$, and $Q$ respectively, where $P, Q$ are continuous on a compact domain, and $d_{wqt}(P,Q) = \frac{1}{2}\int_0^1 (P(Q^{-1}(x)-x)^2 dx$, then
\begin{align}\label{eq:WQTB}
    \frac{1}{n}D_{wqt}(F_n, G_n) \ra_w& \pi_1^2 d_{wqt}(P,Q).
\end{align}
\end{theorem*}
Analogous to Sec.~\ref{sec:WDT}, assuming that the samples live on a compact set, the test statistic is bounded. Thus it follows from the Portmanteau theorem that $\mathbb{E}\big[\frac{1}{n}D_{wqt}(F_n, G_n)\big] \ra \pi_1^2 d_{wqt\mhyphen 1}(P,Q)$, which has the form of \eqref{eq:asymp}. Then, by process described in Sec.~\ref{sec:MF}, the matched filter for the WQT in the operational case when $n$ is finite is $h[t]= \big(1-\frac{|t|}{n}\big)^2$.

Thm.~\ThmWQT~ adds a $\frac{1}{n}$ factor to the definition of the WQT in \eqref{eq:WQTA} that removes a distribution dependent $O(1)$ term in the WQT. In the operational case, we approximate this term by considering the case where $P=Q,~d_{wqt}(P, Q) = 0$, where the $O(1)$ term would have the most impact. This acts as a constant bias term that is removed from the signal prior to matched filter convolution when using the WQT. Details of this can be found in Appx.~\ref{app:WQT}. 

\subsection{Sliced Wasserstein Quantile Test (SWQT)}
Since the WQT is only defined in one dimension, the naive approach for an extension to multiple dimensions is to average the WQT across each dimension independently. Alternatively, we propose to use the sliced Wasserstein quantile test (SWQT) using a similar approach to the sliced Wasserstein distance \cite{bonneel_sliced_2015}, that averages the WQT over one-dimensional projections of the data. Given a two sets of $n$ samples, $\mathcal{F}_n, \mathcal{G}_n$  the SWQT is,
\begin{align}
    D_{swqt}(\mathcal{F}_n, \mathcal{G}_n) \triangleq \ints_{\mathcal{S}^{d-1}} D_{wqt}(F_n^\theta, G_n^\theta) d\theta
\end{align}
where $d\theta$ is the uniform measure on $\mathcal{S}^{d-1}$ - the unit sphere in $\mathbb{R}^d$, and $F_n^\theta$, $G_n^\theta$ are the respective EDFs computed from the projections of the samples on the unit vector $\theta$.

With this definition we state the following,
\begin{theorem*} {\textnormal{\textbf{\ThmSWQT:}}} \label{thm:swqt}
Let sets $\mathcal{F}_n, \mathcal{G}_n$ each consisting of $n$ samples drawn IID from $\left(\pi_1 P + (1-\pi_1) Q \right)$ and $Q$ respectively, where $P, Q$ are continuous on a compact domain, and $d_{swqt}(P,Q) = \int_{\mathcal{S}^{d-1}} d_{wqt}(P^\theta,Q^\theta) d\theta$. Then,
\begin{align}
    \frac{1}{n}D_{swqt}(\mathcal{F}_n, \mathcal{G}_n) \ra_w \pi_1^2 d_{swqt}(P,Q).
\end{align}
\end{theorem*}

Assuming that the samples live on a compact set, the test statistic is bounded. Thus it follows from the Portmanteau theorem that $\mathbb{E}\big[\frac{1}{n}D_{swqt}(F_n, G_n)\big] \ra \pi_1^2 d_{swqt}(P,Q)$, which has the form of \eqref{eq:asymp}. Then, by process described in Sec.~\ref{sec:MF}, the matched filter for the SWQT in the operational case when $n$ is finite is $h[t]= \big(1-\frac{|t|}{n}\big)^2$. 

As with the WQT in Sec.~\ref{sec:WQT}, Thm.~\ThmSWQT~ is stated with a $\frac{1}{n}$ factor that removes an $O(1)$ term. This $O(1)$ term is approximated to by considering the case of $P=Q$ where $d_{swqt}(P,Q)=0$. This acts as a constant bias (and is identical to the bias of the WQT) which is removed from the signal prior to matched filter convolution. Details of this can be found in Appx.~\ref{app:SWQT}.

\subsection{Kolmogorov-Smirnov (KS)}
The two-sample KS test \cite{massey_kolmogorov-smirnov_1951} computes the maximum deviation between the respective empirical distribution functions,
\begin{align} \label{eq:KS}
    D_{KS} (F_n, G_n) \triangleq \sup_x |F_n(x) - G_n(x)|.
\end{align}
Under continuity assumptions on the distributions $F,G$, it is known to be distribution-free under the null hypothesis \cite{pratt_concepts_1981}. 

We note the following theorem for the KS test.

\begin{theorem*} {\textnormal{\textbf{\ThmKS:}}} \label{thm:ks}
Let $F_n$, $G_n$ be derived from IID samples drawn from $\left(\pi_1 P + (1-\pi_1) Q \right)$, and $Q$ respectively, where $P, Q$ are continuous on a compact domain, and $d_{KS}(P,Q) =\sup_x |P(x) - Q(x)|$. Then,
\begin{align}
 D_{KS} (P_n, Q_n) \ra_{as} \pi_1 d_{KS}(P,Q).
\end{align}
\end{theorem*}

Assuming that the samples live on a compact set, the test statistic is bounded. Thus it follows from the  Portmanteau theorem that $\mathbb{E}\left[D_{KS}\left(F_n, G_n\right)\right] \ra \pi_1 d_{KS}(P,Q)$, which has the form of \eqref{eq:asymp}. Then, by process described in Sec.~\ref{sec:MF}, the matched filter for the KS test in the operational case when $n$ is finite is $h[t]= \big(1-\frac{|t|}{n}\big)$.

\subsection{Maximum Mean Discrepancy Squared (MMD$^2$)}
The MMD between two distributions $F,G$, represents the largest difference in expectations over functions in the unit ball of a Reproducing Kernel Hilbert Space (RKHS) with kernel\footnote{We will assume that the RKHS is universal. In this case the MMD is a metric \cite{gretton_kernel_2012}.} $k(\cdot,\cdot)$,
\begin{align}
    \mbox{MMD}(F,G, k) = \sup\limits_{\psi \in \mbox{RKHS}(k): \|\psi\|_k \leq 1} \mathbb{E}_F[\psi] - \mathbb{E}_G[\psi] \nonumber.
\end{align}

In this work we will consider MMD$^2$ statistic for CPD. Given two sets of samples, $\mathcal{F}_n = \{f_1, ... f_n\}$ $\mathcal{G}_n = \{g_1, ... , g_n\}$ sampled IID from $F$ and $G$ respectively, the MMD$^2$ has an unbiased estimator \cite{gretton_kernel_2012} given by,
\begin{align}
    D_{mmd^2}(&\mathcal{F}_n, \mathcal{G}_n, k) = \frac{1}{n^2-n} \sum_{\substack{i,j=1:n \\i\neq j}} \bigg(k(f_i, f_j)  \nonumber \\
    &+  k(g_i, g_j) - k(f_i,g_j) - k(g_i,f_j) \bigg).
\end{align}
Note that under the null hypothesis, the (appropriately scaled) limiting distribution of this unbiased estimator is not distribution-free \cite{gretton_kernel_2012}. We have the following theorem.
\begin{theorem*} {\textnormal{\textbf{\ThmMMD:}}} \label{thm:mmd}
Let sets $\mathcal{F}_n, \mathcal{G}_n$ each consisting of $n$ samples drawn IID from $\left(\pi_1 P + (1-\pi_1) Q \right)$ and $Q$ respectively, where $P, Q$ are continuous on a compact domain, and $d_{mmd^2}(P,Q) = \mathbb{E}_{P \times P}[k(p,p')] + \mathbb{E}_{Q \times Q}[k(q,q')] - 2\mathbb{E}_{P \times Q}[k(p,q')]$ where $p,p'\sim P$ and $q,q'\sim Q$. Then,
\begin{align}\label{eq:mmdU}
    \lim_{n\ra\infty}\mathbb{E}\big[D_{mmd^2}(\mathcal{F}_n, \mathcal{G}_n, k)\big] = \pi_1^2 d_{mmd^2}(P,Q).
\end{align}
\end{theorem*}

Since \eqref{eq:mmdU} has the form of \eqref{eq:asymp}, by the process described in Sec.~\ref{sec:MF}, the matched filter for the MMD$^2$ distance in the operational case when $n$ is finite is $h[t]= \big(1-\frac{|t|}{n}\big)^2$.

\begin{figure*}
\centering
\includegraphics[width=2.0\columnwidth]{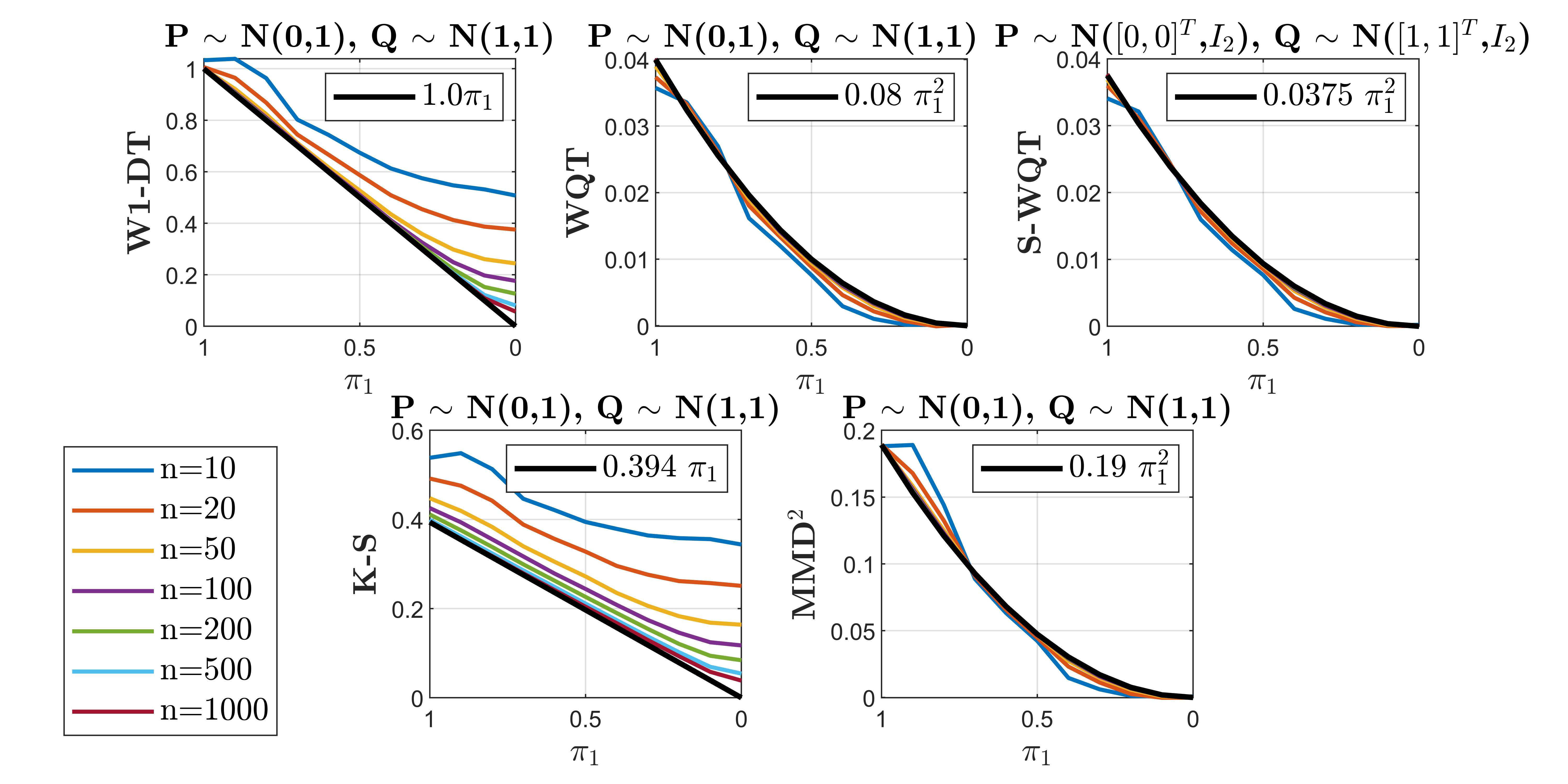}
\caption{Empirical results of simulating the filter for the W1-DT \textbf{(top left)}, WQT scaled by $\frac{1}{n}$ \textbf{(top middle)}, SWQT scaled by $\frac{1}{n}$ \textbf{(top right)}, KS \textbf{(bottom left)} and MMD$^2$ \textbf{(bottom right)} test statistics as a function of mixture parameter $\pi_1$ for various window sizes $n$, along with expected asymptotic result \textbf{(black)}.}
\label{fig:FilterShape}
\end{figure*}

\section{Evaluation}\label{sec:experiments}
The only algorithmic hyperparameters required for the non-parametric statistical change point methods described in this paper are the window size $n$ and the detection threshold parameter $\eta$. For each experiment, we compare filtered and unfiltered version of each statistical test using under the same window parameter.  For all real data experiments, we use domain knowledge of the frequency at which change points should be detected to set the window size. Since not all tests have the same statistical guarantees, we evaluate using metrics that vary the threshold parameter $\eta$ over all possible values.

Early applications of change point detection in failure detection focused on metrics such as average run length and detection delay as key metrics of CPD \cite{tartakovsky_asymptotically_2018, veeravalli_quickest_2012}. Recent applications frame the problem of multiple CPD as a classification problem at each time step (change point vs. no change point) where there is a severe class imbalance as only a small fraction of time steps is regarded as true change points. Here we follow the work of \cite{burg_evaluation_2020} moving toward metrics based on the confusion matrix. We also consider performance over a range of thresholds to allow the end-user maximum control of these trade-offs.
To these ends, when possible, we report the full Precision-Recall (PR) curve, as well as the Area Under the PR Curve (AU-PRC), and best-F1 score (harmonic mean of precision and recall) over across all threshold values.

One motivation for the matched filter approach is to disambiguate multiple local maxima near potential change points. Prior CPD works address this issue by removing from consideration duplicate peaks that are within a pre-specified distance $\delta$ of one another\cite{liu_change-point_2013}, keeping only the highest peak. For a fair comparison to unfiltered methods including prior work, we apply this post-processing to all unfiltered methods, but not the matched filtered methods. For clarity, all methods where duplicate local maxima are removed are labeled with ($\delta$-), and all matched filtered methods where no such post-processing is applied are labeled as ($F$-). 

Thus, for evaluation purposes, we include two additional parameters; $\delta$ is the minimum distance between detected change points applied to the unfiltered methods, and $\epsilon$ defines the tolerated distance for scoring detected change points. Specifically, a detected change point is a True Positive (TP) if there exists a true change point within $\epsilon$ samples, otherwise it is considered a False Positive (FP). False Negatives (FN) are true change points that do not contain any detected change point within $\epsilon$ samples. Precision (P), Recall (R), and F1 are then defined as,
\begin{align}
    P = \frac{TP}{TP+FP},  \hspace{3mm} 
    R = \frac{TP}{TP+FN},  \hspace{3mm}
    F1 = 2\frac{R \cdot P}{R+P}.
\end{align}

\subsection{Simulation Data} \label{sec:simDat}
First, we verify our proposed matched filters with simulated data. Given two known distributions $P$ and $Q$, the exact mixture scenario in Fig.~\ref{fig:diagram} is simulated where $F_n \sim \pi_1 P + (1-\pi_1)Q$ and $G_n \sim Q$.
The test statistic $D_*(F_n(\pi_1), G_n(\pi_1))$ is computed for all values of the mixture parameter $\pi_1$ and for various window sizes $n$. Each test was averaged over 10,000 repetitions using different random seeds. 

Next, we validate the benefits of matched filters for change point detection on simulated sequences using AU-PRC and best-F1 metrics. Separate evaluations are performed for tests defined for scalar and vector valued data. For the scalar case, we generate 40 IID data sequences of length 800 with a single change point uniformly distributed between 300 and 500. Samples prior to the change point are drawn from distribution $P \sim \mathcal{N}(0,1)$, whereas samples after the change point are drawn from $Q \sim \mathcal{N}(0.25, 1)$. 

For the multivariate case, an identical simulation setup is used but $P$ and $Q$ are defined with a shared covariance $\Sigma$ and a difference in mean:
\begin{align}
P &\sim ~\mathcal{N}\bigg(
    \begin{bmatrix}-0.12 \\ +0.12\end{bmatrix},
    \Sigma \bigg),
    \quad
Q \sim ~\mathcal{N}\bigg(
    \begin{bmatrix}
    +0.12 \\ -0.12\end{bmatrix},
    \Sigma \bigg),
    \notag
\end{align}
\begin{align}
\Sigma &= \begin{bmatrix}
    1 & 0.9 \\
    0.9 &1\end{bmatrix}. \nonumber
\end{align}


The SWQT is computed via Monte Carlo simulations by randomly sampling vectors $\theta \sim \mathcal{S}^{d-1}$, and averaging the results over each linear projection. The Gaussian kernel with unit variance is used for computation of the MMD. With these datasets, we compare the performance between the filtered and unfiltered test statistics for various window lengths and evaluate with parameters $\epsilon = \delta = n$.


Finally, the difference in the regions of sensitivity of the Q-Q tests versus the scalability of non-Q-Q tests is illustrated. First we compute the WQT and W1-DT for two uniform distributions, $U[0,1]$ and $U[d,d+1]$ for $d\in [0,2]$, modeling the behavior of these tests for distributions with shifting supports. 
Then, we verify the invariance of the WQT to order-preserving transformations by simulating a time series with a sequence of 4 distributions, $N(0.1, 0.1)$, $N(0.2, 0.4)$, $N(0.4, 0.16)$, $N(0.8,6.4)$ each with 500 samples. We note that by construction, the relative scale at each change point is equal, where each successive segment models the data being scaled by a factor of 2. Thus, if $X$ represents a random variable in the first segment, a random variable in the following three segments would be $2X, 4X, 8X$ respectively. Two additional scenarios are considered; one leaves the data as is, and the other transforms the data by a cubic $X[t]^3$, which is a monotonically increasing, order-preserving function. We then compute the filtered change point statistic using a window of $n=100$ samples comparing the WQT and the W1-DT. The test results were aggregated over 10 independent iterations using AU-PRC as a measure of change point performance.

\subsection{Simulated Results}
\label{sec:SimulatedResults}

\begin{table}[]
    \centering
    \begin{tabular}{c|c|c|c||c|c|c}
          & \multicolumn{3}{c||}{AU-PRC} & \multicolumn{3}{c}{Best-F1}\\
          & n=50 & 100  & 150 & 50 & 100 & 150\\
    \hline
         $\delta$-WQT & 0.52 & 0.76 & 0.90 & 0.46 & 0.69 &0.82 \\
         F-WQT & 0.54 & 0.80 & 0.93 & 0.49 & 0.73 & 0.87\\
         \hline
         $\delta$-MMD$^2$ & 0.47 & 0.75 & 0.88 & 0.45 & 0.67 & 0.83\\
         F-MMD$^2$ & 0.53 & 0.78 & 0.89 & 0.50 & 0.70 & 0.84\\
         \hline
         $\delta$-MW1 & 0.51 & 0.78 & 0.89 & 0.49 & 0.70 & 0.83 \\
         F-MW1 & 0.54 & 0.89 & 0.94 & 0.46 & 0.75 & 0.84 \\
         \hline
         $\delta$-KS & 0.53 & 0.70 & 0.86 & 0.46 & 0.66 & 0.79 \\
         F-KS & 0.54 & 0.88 & 0.98 & 0.46 & 0.72 & 1.0 \\
    \end{tabular}
    \caption{\textnormal{Simulated matched filter results for statistical tests on $\mathbb{R}$ for filtered (denoted by F-) and unfiltered statistics on a series of single change point simulated time series as described in \ref{sec:simDat}. Both AU-PRC and best-F1 scores increase with the inclusion of the matched filter. As expected, performance also improves with increased window length $n$.}}
    \label{tab:filter}
\end{table}

\begin{table}[]
    \centering
    \begin{tabular}{c|c|c|c||c|c|c}
          & \multicolumn{3}{c||}{AU-PRC} & \multicolumn{3}{c}{Best-F1}\\
          & n=50 & 100  & 150 & 50 & 100 & 150\\
    \hline

         $\delta$-MMD$^2$   & 0.19 & 0.67 & 0.85 & 0.36 & 0.65 & 0.88\\
         F-MMD$^2$ & 0.27 & 0.85 & 1.0 & 0.48  & 0.86 & 1.0\\
         \hline
         $\delta$-SWQT & 0.52 & 0.95 & 0.97 & 0.56 & 0.95 & 1.0\\
         F-SWQT & 0.73 &  1.0 & 1.0 & 0.72 & 1.0 & 1.0\\
    \end{tabular}
    \caption{\textnormal{Simulated performance of matched filter on $\mathbb{R}^2$ given the experimental setup described in \ref{sec:simDat}. Both AU-PRC and best-F1 scores increase with the inclusion of matched filtering. SWQT generally outperforms MMD$^2$ for this simulated example.}}
    \label{tab:sliced}
\end{table}

The plots in Fig.~\ref{fig:FilterShape} confirm the results from our theorems and show the convergence of the signature for each of the statistical tests to the expected functional form. Generally speaking, even for sample sizes on the order of $100$ show convergence towards the expected signature.

In the simulated change point tests on $\mathbb{R}^1$ (Tab.~\ref{tab:filter}), and $\mathbb{R}^2$ (Tab.~\ref{tab:sliced}) show that the application of our proposed matched filters to the corresponding test statistic yields consistent improvement in the AU-PRC and best-F1 metrics thus resulting in an improved true positive to false positive ratio. As expected, when window size increases, overall detection performance also increases. In this controlled setting, the performance across all four possible statistical tests is comparable in the univariate case. However, in the two-dimensional case the SWQT has a better overall performance compared to the MMD$^2$ across all window sizes.
\begin{figure}
\centering
\raisebox{\dimexpr 1.0\baselineskip-\height}
{\includegraphics[width=0.4\textwidth]{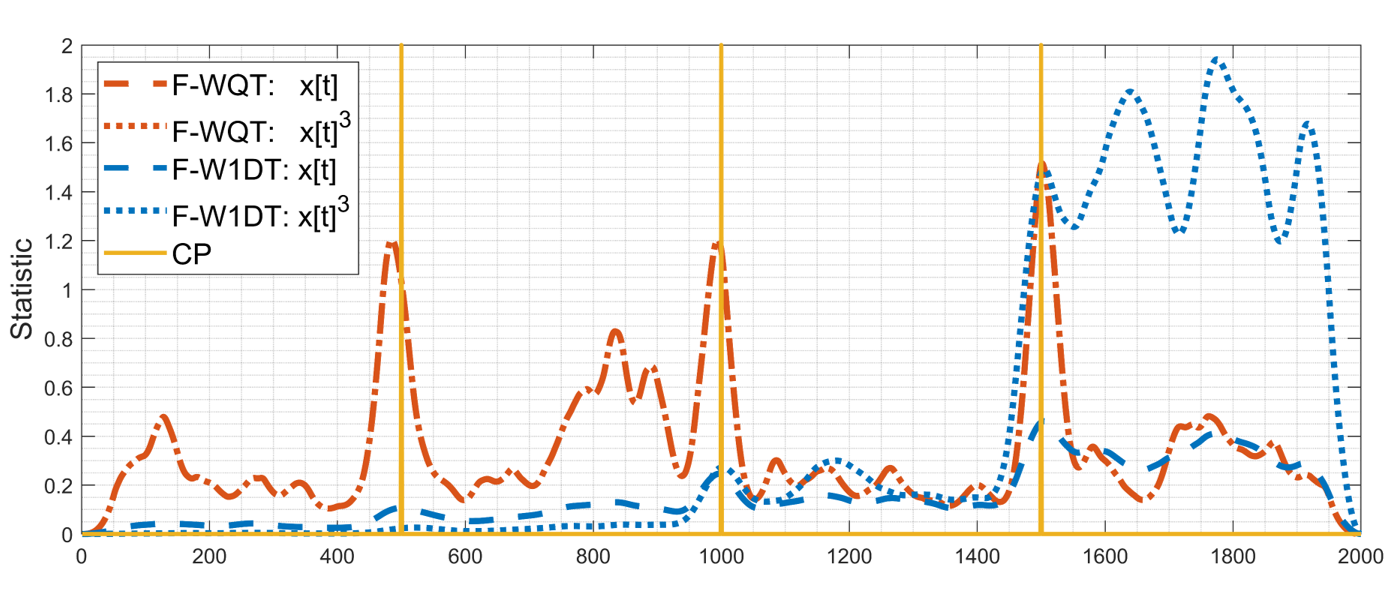}}\hfil
\\
\medskip
\begin{tabular}[t]{c|cc}
\multicolumn{3}{c}{}\\
AU-PRC&$X[t]$ & $X[t]^3$ \\
\hline
F-WQT &0.865 & 0.846\\
F-W1DT &0.349 & 0.254 \\
\end{tabular}
\raisebox{\dimexpr 0.6\baselineskip-\height}
{\includegraphics[width=0.25\textwidth]{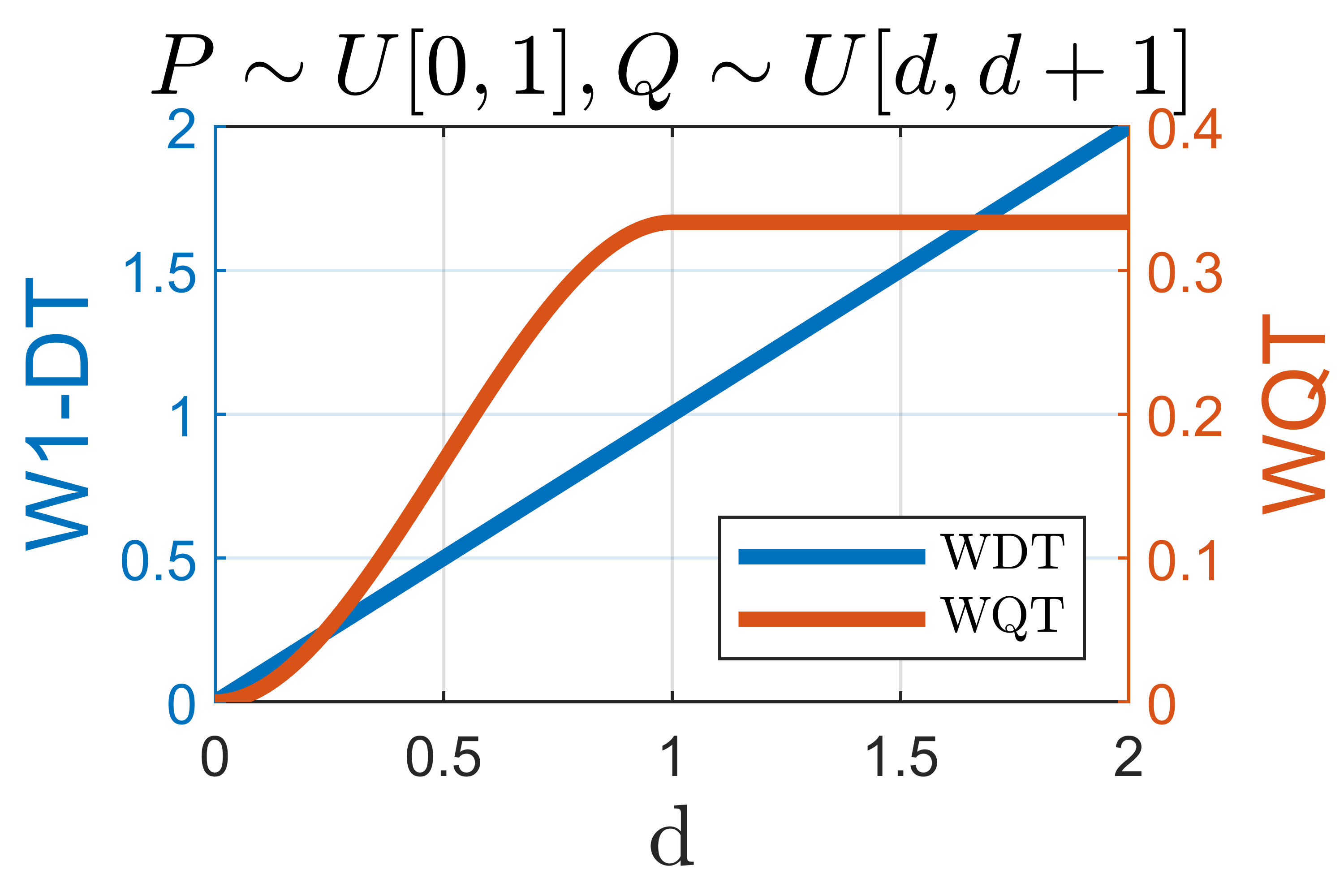}}\hfil

\caption{Simulated CPD comparing the W1-DT (EDF/QF test) with the WQT (Q-Q test). Sample output \textbf{(top)} of matched filtered WQT and W1-DT output for simulated data with and without cubic transformation. We note the two traces for the F-WQT are coincident. AU-PRC values \textbf{(bottom left)} averaged over 10 runs. Clearly, as all 3 change points are labeled as true changes, the WQT vastly out-performs the W1-DT. Comparison of WQT and W1-DT \textbf{(bottom right)} on uniform distributions with shifting supports computed based on their definitions in Sec.~\ref{sec:WDT}, Sec.~\ref{sec:WQT} exhibiting the differences in sensitivity of the two tests in different regions.}\label{fig:WdtWqt}
\end{figure}

Other commonly used CPD evaluation metrics consider detection error between the true and labeled change points (e.g. mean absolute error, or mean squared error \cite{aminikhanghahi_survey_2017}). In our simulated and real-world experiments, we found no significant improvement in detection error-based metrics when applying the matched filter. For example, in this single change point simulated context, for WQT $n=150$, the MAE for the unfiltered and filtered is $46.8$ and $45.5$ samples respectively.

Since the data matches all of our assumptions, we are also able to empirically verify the peak-preserving property of the matched filter as shown in Fig.~\ref{fig:motivation}, where the data is generated by sampling IID alternating between two normal distributions $\mathcal{N}(0,1), \mathcal{N}(0.25,1)$ with a window size of $n=150$.


The difference between the WQT (Q-Q based), with the W1-DT (EDF, QF based), for CPD are highlighted using simulated data in Fig.~\ref{fig:WdtWqt}. In the simulated time series, at each successive segment where the mean and standard deviation of the data is doubled, the WQT detects each change with essentially the same magnitude, thus a single threshold suffices to detect these three change points as the relative change is constant. Conversely, the change point response of the W1-DT scales with the absolute magnitude of the change in the signal. Furthermore, since the Q-Q is invariant to order-preserving transformations, the change point statistic of the WQT under the cubic transform on the data remains identical but the W1-DT statistics shifts drastically. 

When comparing the WQT and W1-DT on uniform distributions of shifting supports, Fig.~\ref{fig:WdtWqt} shows both the bounded property of the WQT, saturating when the supports of the two distributions become disjoint ($d>1$), and the difference in sensitivity (slope of the response) of the two tests. Straightforward calculations show that the W1-DT is equally sensitive (constant slope) regardless of the shift in $d$. In contrast, the WQT shows different regions of sensitivity.  Specifically, the WQT it is more sensitive to small changes in support but is insensitive to any additional change once the supports are disjoint. While the scales of these two tests differ as evident from the left and right axes labels, all evaluation in this paper is based on local maxima and precision-recall curves the structure of which is independent of the absolute amplitude of the test statistic.  We return to this point in the discussion of the real-world data below.

In summary, two properties of Q-Q tests for CPD are (1) the ability to use a single threshold to detect changes at different scales, and (2) the high sensitivity of these tests to small changes in data support. In some applications where these characteristics are desirable, change point methods built on Q-Q tests will provide better results as seen from the clear benefits in AU-PRC in Fig.~\ref{fig:WdtWqt}. However, in cases where perhaps the absolute magnitude of the change significant, tests based on the EDF or QF would produce better results. For comparison, the reported results in the table of Fig.~\ref{fig:WdtWqt} uses all three change points as true change points. If only the change with the greatest magnitude (that is, the third one) was considered a ``true'' change, the performance would be reverse with an AU-PRC of 0.256 for the F-WQT and 0.687 for F-W1DT.

\begin{figure*}
    \centering
\includegraphics[width=2.1\columnwidth]{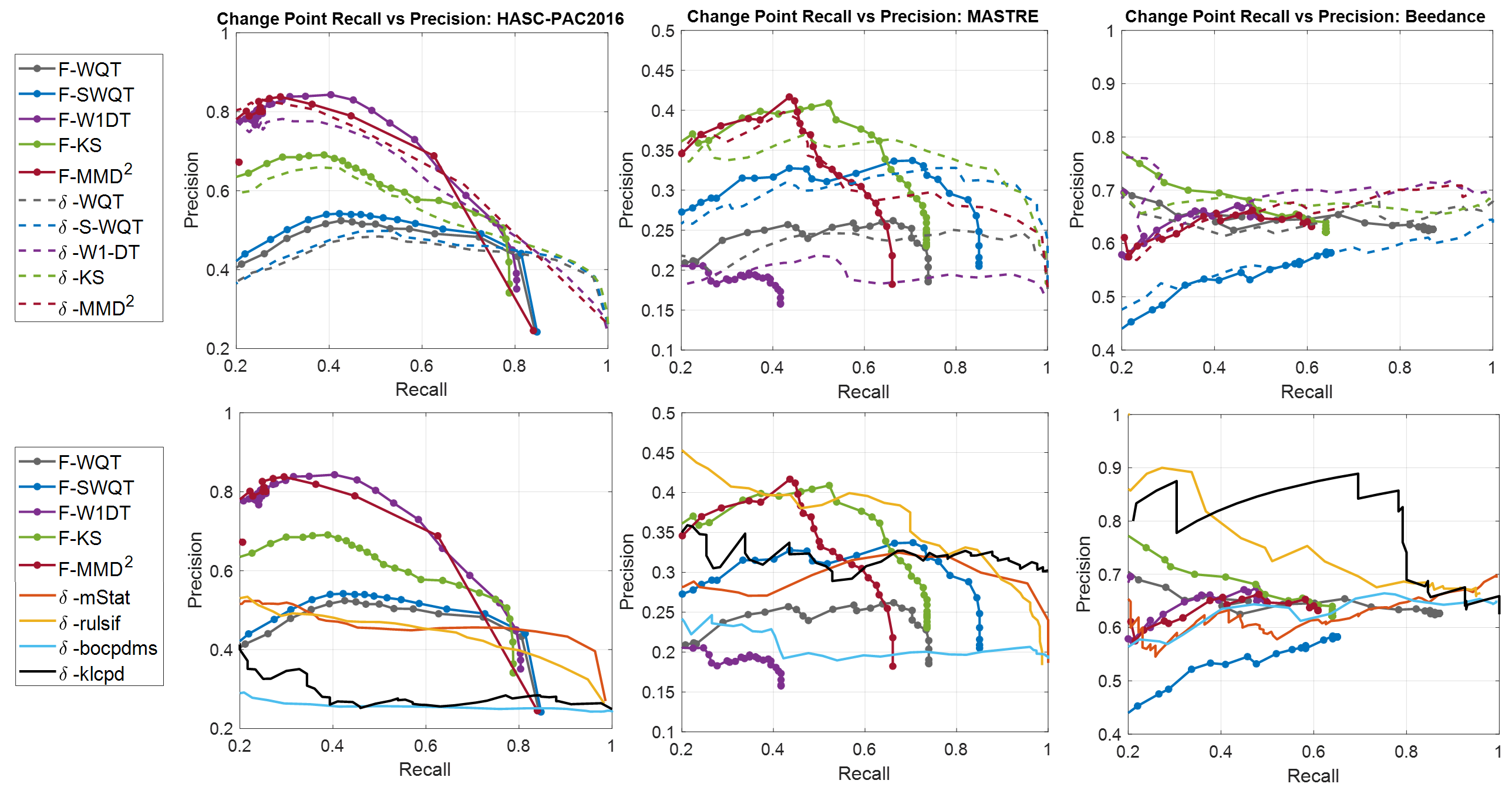}
    \caption{Precision vs recall evaluated on HASC (\textbf{left column}, window size $n=200$), MASTRE (\textbf{middle column}, $n=200$), and Beedance (\textbf{right column}, $n=20$) benchmark tasks.
    For each dataset we compare \textbf{(top row)}, unfiltered (dashed) and matched filtered (solid) statistics using SWQT, WDT, MMD$^2$, and KS statistical tests. We also compare \textbf{(bottom row)} the matched filtered statistic with results from using the M-Statistic, RulSIF, BOCPD-MS, and KL-CPD (supervised).
    Our proposed matched filtered versions (denoted with prefix ``$F$-'') do not require duplicate detection post-processing.
    The unfiltered versions (denoted with prefix ``$\delta-$'') are post-processed to remove any detections within $\delta$ samples (HASC $\delta=150$, MASTRE $\delta=150$, Beedance $\delta=16$).
    }
    \label{fig:prAll}
\end{figure*}

\subsection{Real-World Data}
\label{sec:RealWorldData}
We compare the filtered and unfiltered versions of the statistical tests described in this paper with prior work using identical windowing parameters where applicable. The M-statistic \cite{liu_change-point_2013} is a sliding window CPD algorithm based on the MMD. BOCPD-MS \cite{knoblauch_spatio-temporal_nodate}, is a parametric Bayesian method that extends \cite{adams_bayesian_2007} through model selection. Here we utilize the algorithms default parameters\footnote{Code from \url{https://github.com/alan-turing-institute/bocpdms}.} and set the change point statistic as the log probability that the run length equals zero. RuLSIF \cite{liu_change-point_2013} uses direct density ratio estimates between sliding windows\footnote{Code from \url{https://riken-yamada.github.io/RuLSIF.html}.}. KL-CPD \cite{chang_kernel_2019} applies the MMD on sliding windows with a kernel trained as a neural network in a supervised setting. We note that KL-CPD is the only supervised method included in our evaluation. In this supervised setting, since data is required for training and validation, KL-CPD is tested on a subset of the available evaluation data. The default setup\footnote{Code from \url{https://github.com/OctoberChang/klcpd_code}.} is used where 60\% of the each sequence is used for training, 20\% for validation, and 20\% for testing. Therefore, comparison of KL-CPD to all other unsupervised methods should be considered carefully. To each of these methods, we remove duplicate peaks within $\delta$ samples of each other.

To provide a detailed analysis of performance, we plot the PR curve for each method evaluated on the following datasets:

\textbf{HASC-PAC2016} \cite{ichino_hasc-pac2016:_2016} $n = 200, \delta=150, \epsilon=150$: A raw dataset that consists of over 700 three-axis accelerometer sequences sampled at 100 Hz of subjects performing six actions: 'stay', 'walk', 'jog', 'skip', 'stairs up', and 'stairs down'. The 92 longest sequences where each of the six actions are represented are used for evaluation. Time series have an average length of 17,775 samples and 15.2 change points. 

\textbf{MASTRE}\cite{hussey_monitoring_2020} $n=200, \delta=150, \epsilon=150$:
In this proprietary dataset, soldiers move between a series of stations to perform various physical tasks.
The nature of the tasks varies from marksmanship to aerobic exercises.
Subjects are instrumented with a three-axis accelerometer sampled at 100 Hz, and change points are labeled from video as the subject transitions into and out of tasks.
A total of three time sequences were evaluated with an average length of 92,097 samples and 65.3 change points

\textbf{Beedance} \cite{oh_learning_2008} $n~= 20$, $\delta=16$, $\epsilon=16$: A dataset containing movements of dancing honeybees who communicate through three actions: "turn left", "turn right" and "waggle". 
A total of 6 time sequences are evaluated, each one containing 3-dimensional signal of the X,Y location and heading angle of the bee as captured in an overhead image. The time series have on average a length of 787 samples, and 18.8 change points. We obtained the positions and angles from the original data release.


For datasets with multiple dimensions, methods inherently defined for univariate signals ($X[t] \in \mathbb{R}$) are extended to higher dimensions by averaging their respective test statistic over each dimension. This applies to WQT, W1-DT, KS, and RuLSIF.

\subsection{Real-World Results}

For HASC dataset, where the window size is large ($n=200$), the left column of Fig.~\ref{fig:prAll}  shows the improvement in performance provided by the matched filter. At each possible fixed recall, there is consistently about a 5\% increase in precision with the matched filter applied up to a recall of about $0.78$, indicating that in this region, the matched filter decreases false positives without increasing false negatives.  However, recall values past a certain point are not achievable by the matched filtered methods under the given algorithm parameter $n$ and evaluation parameter $\epsilon$. This is especially true as the threshold $\eta$ becomes small. In this regime, the stochastic nature of the unfiltered test statistics often produces what are essentially spurious, local maxima \textit{someplace} within $\pm \epsilon$ of a true change point.  The matched filter however serves as a low-pass filter, removing these peaks resulting in recall values less than unity.  In these cases, the higher achievable recall of the unfiltered statistical tests is attributed to the prevalence of local maxima due to randomness rather than the inherent properties of the test.
Interestingly, the best-F1 scores are comparable between the filtered and unfiltered methods in the HASC dataset, and are achieved around where the two curves cross.  Thus, up to a certain recall, the application of matched filtering improves detection precision and F1 score. Past this recall, only achievable by unfiltered methods, the use of matched filtering does not provide benefits in terms of F1 score. 

For MASTRE, the performance of matched filtered statistics in Fig.~\ref{fig:prAll} (middle column) compared to unfiltered methods shows similar trends to HASC where for lower recall values, matched filtered statistics consistently produce higher precision values. However unlike HASC, the recall value at which the filtered methods fall off differ significantly between each of the test statistics. This discrepancy is due to the difference in how true change points are labeled between the two datasets, discussed in depth below.

Referring to the right column in Fig.~\ref{fig:prAll}, for the Beedance dataset, the matched filter does not seem to offer clear benefits compared to the baseline test statistic signals.
In fact, we might expect this because the small window size required for this dataset ($n=20$) likely means we are far from the asymptotic regime of $n \rightarrow \infty$ in which the matched filters are derived.
We thus include this result as a known limitation as we expect matched filters to show benefits only for large window sizes.

As seen in the bottom row of Fig.~\ref{fig:prAll}, compared to prior work, the relative performance of sliding window statistical tests discussed in this work varies depending on the dataset. For the HASC data, all statistical tests show better results compared to all other evaluated methods. In the cases of MASTRE, RuLSIF shows the overall best results. Notably, mStat performs comparably to the unfiltered SWQT for both HASC and MASTRE.  For Beedance, KL-CPD (supervised) shows the best results. Of the prior methods evaluated, RuLSIF performs most consistently over the three datasets.

Furthermore, we note that for the MASTRE data set, there is a significant improvement in the SWQT compared to the WQT averaged across each dimension whereas for HASC, there is only a slight improvement. From this we can deduce that the change points of HASC can be observed through the analysis of each dimension independently, whereas for MASTRE performance is improved by considering vector valued methods.

\begin{figure*}
\centering
\includegraphics[width=2.0\columnwidth]{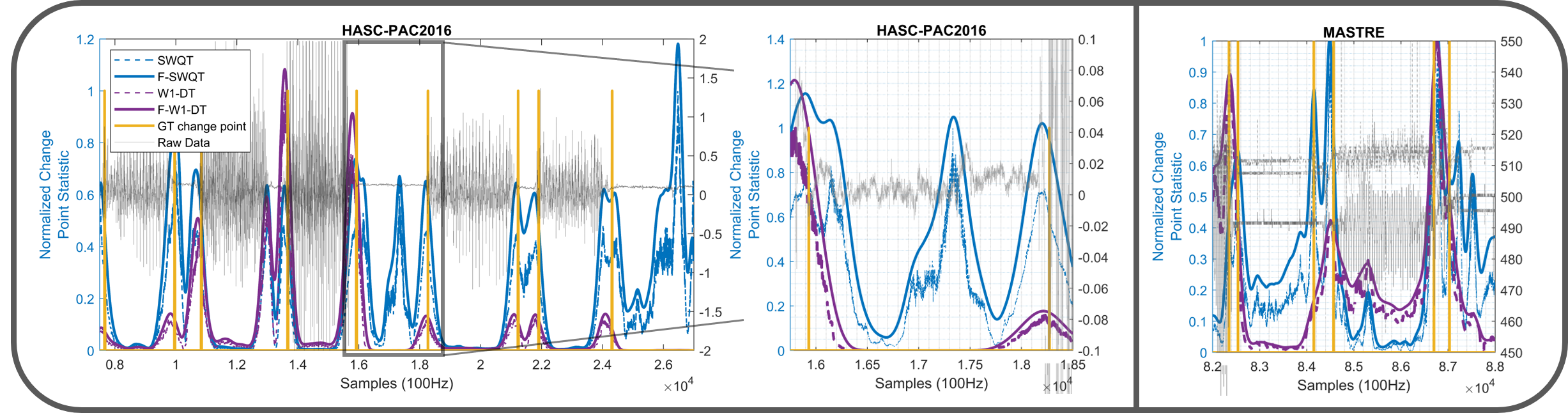}
\caption{Sample output for HASC-PAC2016 \textbf{(left, middle)} and MASTRE \textbf{(right)} human activity accelerometer data sequence (grey) and the ground truth change points (yellow), with the filtered (solid) and unfiltered (dashed) SWQT (blue), and W1-DT (purple). For comparison, each statistic is normalized based on the maximum value of their respective unfiltered statistic over the entire sequence. While it appears that the SWQT false alarms frequently \textbf{(left)}, zooming into one such region \textbf{(middle)} shows a small shift in support resulting in a change in the data at a smaller scale which HASC does not consider a true change point. The example sequence from the MASTRE data (right) shows a similar small shift that is a true change point that are detected by the SWQT but not the W1-DT.}
\label{fig:hascW2tOt}
\end{figure*}

The HASC and MASTRE datasets represent human activity measured through accelerometry under different contexts. The HASC experiment measures human activity in a controlled setting where subjects are instructed to hold an action until switching to another one of the six allowed actions. On the other hand, MASTRE is collected in a task-oriented setting where transitions between individual actions are more fluid, which is one reason why overall performance on MASTRE is lower. Furthermore, while there are running and walking tasks similar to that of HASC, the MASTRE data also encompass changes where the individual is not moving their feet, such as standing-to-kneeling posture changes. 

The characteristics of the HASC and MASTRE data sets noted in the previous paragraph lead to differences in performance among the tests.  In the HASC precision-recall curve, the performance of the statistical tests varies. Given a certain recall value, W1-DT and MMD$^2$ have the highest precision, KS is in the middle, and the WQT and SWQT generally have the lowest precision.  The WQT and SWQT results arise from the fact that the Q-Q tests on the HASC data false alarm more often compare to the other tests.  Although the WQT and SWQT achieve a slightly higher recall (for example at a precision level of 0.45), this benefit does not outweigh the loss in precision. However, for the MASTRE PR curves, the story is different. For low recall values (0.2-0.6) there is a similar trend where the MMD$^2$ and KS have higher precision values than the SWQT. However, the discrepancy between the recall values (for example at a precision level of 0.3) is much more pronounced where SWQT achieves a recall significantly higher than that of the other statistical tests, especially the W1-DT which overall performs very poorly on this dataset. These differences in performance of the W1-DT and the WQT/SWQT between HASC and MASTRE can be explained by two factors: (1) the properties of Q-Q tests as discussed in the context of simulated data and (2) how true labels are assigned in the respective datasets. 

As seen from the sample HASC time series (Fig.~\ref{fig:hascW2tOt} left) the WQT generally has peaks of equal height whereas the peaks of the W1-DT scale with the observed magnitude of changes.  This behavior is consistent with the discussion surrounding Fig.~\ref{fig:WdtWqt} concerning the manner in which these tests respond to changes that vary in scale.  Furthermore, while it appears that the WQT false alarms in the stationary regions where the subject is motionless, closer inspection into one of these regions (Fig.~\ref{fig:hascW2tOt}, middle) shows that the signal has a slight shift in mean perhaps due to a shift in posture.  As discussed above (Fig.~\ref{fig:WdtWqt}), the WQT is highly sensitive to these small change in the support of the data resulting in a clear peak in the Q-Q test statistic in Fig.~\ref{fig:hascW2tOt}. This change in the data is small relative to others across the full time series. Thus, consistent with the relative \textit{insensitivity} of EDF/QF tests, it is \textit{not} reflected in the W1-DT. The ground truth change point labels of HASC focuses more on the large-scale activity changes, therefore change points are not labeled for these small changes, which contributes to the poor precision of the Q-Q tests. 

Change points in the MASTRE dataset correspond to entering and exiting stations where tasks are performed, not necessarily based on specific action the subject is taking. Therefore, true change point labels correspond to both large changes in action and subtle changes in posture (Fig.~\ref{fig:hascW2tOt}, right); the latter of which would not be labeled as a true change point in HASC setting. As shown earlier and seen in Fig.~\ref{fig:WdtWqt}, the WQT is sensitive to both these changes, and therefore achieves a higher precision and recall. 

These two human activity datasets provide an example of how the application dictates the suitability Q-Q tests. We have shown how the WQT is particularly sensitive to small changes and support, and that Q-Q tests equally detect changes at different scales. In applications where these properties reflect true change points, as is in the case of MASTRE, Q-Q tests will yield better results. However, in applications similar to the HASC dataset, where subtle shifts in posture are considered false alarms, the W1-DT would be preferred as they would be dwarfed by the larger changes in the time series.

\section{Conclusion and Future Work}
While many methods of change point detection have been proposed over the years, the issue of change point localization for a noisy distribution-free statistic has not been thoroughly considered. To address this issue, we introduce asymptotically matched filters. For various non-parametric tests that have been used as the foundation of multiple CPD algorithms, we derive these filters under the simple observation that sliding windows over a change point will cause samples from one window to be drawn from a mixture distribution. With asymptotic analysis, we are able to derive the expected response of the test statistic in the region of a change point which is then used to compute the matched filter in the operational (non-asymptotic) case. While in this paper we only consider a subset of tests, the proposed analysis methodology for deriving matched filters can be applied to other methods in change point detection.

The discussed framework for change point detection through the use of a two-sample test of sliding windows is both simple and easily deployed in practice. Once a test statistic is chosen, the only hyperparameters required are the window size and detection threshold. We build on this methodology by applying matched filtering which results in improved change point precision, and also simplifies the of the process of identifying change points given a statistic, removing any need for ad-hoc processing to remove duplicate peaks. While similar smoothing effects can be achieved from other low-pass filters, these filters are not guaranteed to be distribution-free or peak-preserving which are properties of matched filters that ensure that statistical guarantees are preserved for statistics where \eqref{eq:asymp} applies. Furthermore, if the test statistic is itself distribution-free, these guarantees are preserved with a constant threshold. While simple, this method of detecting changes points by testing for changes in distribution through two sample tests demonstrates competitive performance with other state-of-the-art approaches.

In understanding the trade-offs between various CPD methods, we build on two properties of Q-Q based statistical tests; namely their invariance to order-preserving transformations and their sensitivity especially to small changes in support of the data (or equivalently, small changes in the mean). For CPD applications these properties result in differences in response.  Specifically, Q-Q tests can detect changes at different scales of the data using a single threshold while tests based on quantile functions or empirical distribution functions tend to be ``tuned'' to changes of a specific magnitude.  As evidenced by our real-world data examples, these differences can be leveraged to properly select the appropriate test for an application, and certainly motivate further rigorous investigation.

Despite the fact that the derivations for the filters in this paper assume that the data is IID, based on real-world results, we see that the benefits still hold on non-IID data when the window size is sufficiently large. Nonetheless, in future work, we hope to consider analysis under non-IID conditions and evaluate if matched filters can be applied to other change point methods. Furthermore, while the primary application of our work focuses on activity data in an offline setting, it is amenable to online applications where some delay can be tolerated. To this end, future work would be needed both in terms of simulation analysis and a sequential mathematical framework for quantifying the performance of our methods in terms of common online CPD metrics such as average run length, and detection delay. Additional applications of this work can be explored in both offline and online contexts with similar data models including but not limited to sensor networks \cite{ciuonzo_generalized_2017}, and sonar systems \cite{myers_synthetic_2020}.

\bibliographystyle{ieeetr}
\bibliography{References}

\begin{appendix}
\subsection{Mathematical Background}\label{sec:background}
The empirical distribution functions, quantile functions and quantile-quantile functions \eqref{eq:EDF_QF_QQ_1}, \eqref{eq:EDF_QF_QQ_2} discussed in this work are each c\`{a}dl\`{a}g functions (right continuous functions, with left hand limits) over their respective domains. Equipped with the Skorohod metric, the set $\mathcal{X}$ of all c\`{a}dl\`{a}g functions on a compact domain is a separable metric space. Let $P_n, P$ be probability measures belonging to $(\mathcal{X}, \mathcal{S})$ where $\mathcal{S}$ represents the Borel $\sigma$-field generated by open sets. 

Weak convergence on metric spaces is defined as,

\begin{definition*}\cite{van_der_vaart_weak_1996}
$P_n \ra_w P$ iff $\int_{\mathcal{X}} f\,dP_n \ra \int_{\mathcal{X}} f\,dP$ for all $f\in C(\mathcal{X})$, where $C(\mathcal{X})$ denotes the set of all bounded, continuous, real functions on $\mathcal{X}$.
\end{definition*}

To understand the behavior of stochastic processes under mappings, we utilize the continuous mapping,  Slutsky's, and the Portmanteau theorems stated below.

\begin{theorem*}~{\textnormal{\textbf{\ThmCM:}}}~\textbf{Continuous Mapping Theorem:}\label{thm:cm}
(2.7 from \cite{billingsley_convergence_1999}),   Suppose there exists a sequence of measures $P_n$ and $P$ that belong $(\mathcal{X}, \mathcal{S})$. Given a mapping $h$ from $\mathcal{X}$ to $\mathcal{X}'$ such that $h(P_n), h(P) \in \mathcal{X}'$, and $D_h$ is the set of discontinuities of $h$ in $\mathcal{X}$, if $P_n \ra_w P$, and $PD_h=0$, then $P_nh^{-1} \ra_w P h^{-1}$
\end{theorem*}

\begin{theorem*}~{\textnormal{\textbf{\ThmSlutsky:}}}~\textbf{Slutsky's theorem:} \label{thm:slutsky} (1.4.7 from \cite{van_der_vaart_weak_1996}), 
If $P_n \ra_w P$ and $Y_n \ra_w c$ where $X \in \mathcal{X}$ is separable, and $c$ is a constant, then $(P_n, Y_n) \ra_w(X,c)$. Furthermore, if $\mathcal{X}$ is a topological vector space then, 
\begin{itemize}[noitemsep]
    \item[(i)] $P_n + Y_n \rightarrow_w P+ c$
    \item[(ii)] $P_nY_n \rightarrow_w cP$
    \item[(iii)] $P_n/Y_n \rightarrow_w X/c\hspace{5mm}$    provided $c\neq 0$.
\end{itemize}
\end{theorem*}

\begin{theorem*}~{\textnormal{\textbf{\ThmPort:}}}~\label{thm:port}\textbf{Portmanteau theorem:}
(2.1 from \cite{billingsley_convergence_1999}) For probability measures $P_n$, $P$ on $(\mathcal{X}, \mathcal{S}))$, the following are equivalent:\footnote{We only state 2 of the 5 equivalent statements of the theorem} \begin{itemize}[noitemsep]
    \item[(i)] $X_n \ra_w P$
    \item[(ii)] $\mathbb{E}[f(X_n)]\ra \mathbb{E}[f(X)]$ for all bounded uniformly continuous $f$
   
\end{itemize}
\end{theorem*}

We state the Glivenko-Cantelli theorem often used in conjunction with the Kolmogorov-Smirnov test. Here, $\ra_{as}$ denotes \textit{almost-sure convergence} which is a stronger condition than weak convergence.
\begin{theorem*}~\label{thm:gc}{\textnormal{\textbf{\ThmGC:}}}~\textbf{Glivenko-Cantelli Theorem:}~(19.1 from \cite{vaart_asymptotic_1998}), 
If $X_1, X_2, \dots$ are IID random variables with distribution function F, then $||F_n-F||_\infty \ra_{as} 0$.
\end{theorem*}

Finally, we state some recent results for the Wasserstein quantile test.

\begin{theorem*}~\label{thm:ramdas}{\textnormal{\textbf{\ThmRamdas:}}}~(From \cite{ramdas_wasserstein_2015}) 
For CDF's F, G with respective densities f,g such that $\frac{g(F^{-1}(x))}{f(F^{-1}(x))} \leq C$, for all $x \in [0,1]$, and for empirical distributions $F_n$ and $G_m$ of $n,m$ samples respectively, where $\frac{n}{m} =\lambda \in [0,\infty)$ as $n,m \ra \infty$,
\begin{align} \label{eq:Ramdas1}
     &\sqrt{\frac{nm}{n+m}}\left(G_m(F_n^{-1}(\cdot))-G(F^{-1}(\cdot)) \right) \nonumber\\ 
     &\hspace{5mm}\ra_w  \sqrt{\frac{\lambda}{\lambda+1}} \mathcal{B}_1(G\cdot F^{-1}(\cdot)) + \sqrt{\frac{1}{1+\lambda}}\frac{g(F^{-1}(\cdot))}{f(F^{-1}(\cdot))} \mathcal{B}_2(\cdot)
\end{align}
where $\mathcal{B}_1(x), \mathcal{B}_2(x)$ are independent Brownian bridges on the interval $[0,1]$.
\end{theorem*}

With this background we prove the asymptotic results for our various statistical two-sample tests.

\subsection{Proof of Theorem \hyperref[thm:wdt]{\ThmWDT}: Wasserstein-1 Distance in $\mathbb{R}$}\label{app:W1-DT}


\begin{proof}
The W1-DT can be expressed as,
\begin{align}\label{eq:wdt_def}
    D_{MW1}(F_n, G_n) = \int_{\mathcal{C}} |\underbrace{F_n(x) - G_n(x)}_{A_n(x)}|dx.
\end{align}

\noindent By Glivenko-Cantelli,
\begin{align}
    \sup_x |F_n(x) - F(x)| &\ra_{as} 0 \\
    \sup_x |G_n(x) - G(x)| &\ra_{as} 0.
\end{align}

\noindent With the triangle inequality,
\begin{align}
    0 &\leq \sup_x |(F_n(x) - F(x)) - (G_n(x) - G(x))|  \\
    & = \sup_x |A_n(x) - (F(x)-G(x))| \\
     &\leq \sup_x |\left(F_n(x) - F(x)\right)| + |\left(G_n(x) - G(x)\right)| \ra_{as} 0.
\end{align}
Therefore, 
\begin{align}
    A_n(x) &\ra_{as} (F(x)-G(x))
\end{align}

Since almost sure convergence implies weak convergence, we can apply the continuous mapping theorem for the function $\int_{\mathcal{C}} f(x) dx$ \cite{mikosch_lecture_2005} (Thm.~\hyperref[thm:cm]{\ThmCM})
\begin{align} \label{eq:wdt_end}
     D_{MW1}(F_n, G_m) &= \int_{\mathcal{C}}|A_n(x)| dx   \nonumber \\
     &\ra_w  \pi_1\int_{\mathcal{C}} |(P(x)-Q(x))| dx \nonumber \\
     &= \pi_1 d_{w1\mhyphen dt}(P,Q).
\end{align}
\end{proof}

Since $F_n, G_n$ map from $\mathcal{C}\ra [0,1]$, the distance \eqref{eq:wdt_def} is also bounded. By Thm.~\hyperref[thm:wdt]{\ThmWDT}~and the Portmaneau theorem, it follows that, 
\begin{align}
    \mathbb{E}\left[D_{w1\mhyphen dt}(F_n,G_n)\right] \ra_w \pi_1 d_{w1\mhyphen dt}(P,Q).
\end{align}
Lastly, we see that the W1-DT is symmetric since $|F_n(x) - G_n(x)| = |G_n(x) - F_n(x)|$.

\subsection{Proof of Theorem \hyperref[thm:wqt]{\ThmWQT}: Wasserstein Quantile Test} \label{app:WQT}
\begin{proof}
By definition of the WQT \cite{ramdas_wasserstein_2015},
\begin{align} \label{eq:wqt}
    \frac{1}{n}D_{wqt} (F_n, G_n) =& \frac{1}{2}\int_0^1(F_n(G_n^{-1}(x))-x)^2 dx \nonumber \\
    =& \frac{1}{2}\int_0^1\Bigg(\underbrace{\left(F_n(G_m^{-1}(x)) - F(G^{-1}(x))\right)}_{\text{$A_n(x)$}} \nonumber \\
    & \hspace{8mm} + \underbrace{\left(F(G^{-1}(x))-x\right)}_{\text{$B(x)$}}\Bigg)^2 dx.
\end{align}
For a given $P$ and $Q$, $B(x)$ is deterministic. To $A_n$ we apply Thm.~\hyperref[thm:ramdas]{\ThmRamdas},
\begin{align} \label{eq:Ax}
    \sqrt{\frac{n}{2}}A_n(\cdot) \ra_w \sqrt{\frac{1}{2}} \mathcal{B}_1(G\circ F^{-1}(\cdot)) + \sqrt{\frac{1}{2}}\frac{g(F^{-1}(\cdot))}{f(F^{-1}(\cdot))} \mathcal{B}_2(\cdot).
\end{align}
Given that the sequence $\sqrt{\frac{2}{n}} \ra 0$, we apply Slutsky's theorem to \eqref{eq:Ax},
\begin{align}
    A_n &\ra_w 0_n
\end{align}
\noindent where $0_n$ denotes a function that is 0 over its domain.

Therefore, since $F=\pi_1 P + (1-\pi_1) Q$, and $G=Q$,
\begin{align}
    A_n + B &\ra_w F(G^{-1}(x)) - x  \label{eq:sym} \\
    (A_n(x) + B(x))^2 &\ra_w \left(\left(\left(\pi_1P + \left(1-\pi_1\right)Q\right)\circ Q^{-1}\right)\left(x\right) - x\right)^2 \nonumber \\
    &=\bigg(\pi_1P\left(Q^{-1}\left(x\right)\right) + Q\left(Q^{-1}\left(x\right)\right) \nonumber\\
    &\tab -\pi_1Q\left(Q^{-1}\left(x\right)\right) - x\bigg)^2\nonumber \\
    &=\pi_1^2\left(P\left(Q^{-1}(x)\right) - x\right)^2
\end{align}
Therefore, by the continuous mapping theorem for the continuous function $\int_0^1 f(x) dx$ \cite{mikosch_lecture_2005},
\begin{align} \label{eq:wqt_cm}
    \frac{1}{n}D_{wqt}(F_n,G_n) &\ra_w \frac{\pi_1^2}{2}\ints_0^1 \left(P\left(Q^{-1}(x)\right) - x\right)^2dx \nonumber\\
    &= \pi_1^2 d_{wqt}(P,Q).
\end{align}
\end{proof}

Once again, because $F_n, G_n: \mathcal{C}\ra [0,1]$, the WQT is bounded. Thus, by Thm.~\hyperref[thm:wqt]{\ThmWQT}~and the Portmanteau Theorem, it follows that, 
\begin{align}
    \mathbb{E}\left[\frac{1}{n}D_{wqt}(F_n,G_n)\right] \ra \pi_1^2 d_{wqt}(P,Q).
\end{align}

In this proof we used Slutsky's theorem for the sequence $\sqrt{\frac{2}{n}}$ to asymptotically remove an $O(1)$ term $A_n$. This suggests that for a finite window size $n$, $\mathbb{E}\left[D_{wqt}(F_n, G_n)\right] = n\pi_1^2 d_{wqt} + O(1)$. We note that this $O(1)$ is \textit{not} distribution-free, but is bounded given that the conditions of the theorem are met. However, we note that this term only becomes a significant factor when $n$ and $d_{wqt}$ are small.

Operationally, we estimate the $O(1)$ term by considering where $d_{wqt}(P,Q)=0$ (or $P=Q$), since it is under this condition where its effect is most prominent. This case is identical to the null hypothesis of the WQT.  \cite{ramdas_wasserstein_2015} shows that under the null hypothesis that $P=Q$, $D_{wqt}(P_n,Q_n)\ra_w \int_0^1 \BB(x)^2 dx$, where $\BB$ is a Brownian bridge on $[0,1]$, and \cite{tolmatz_distribution_2002} shows that $\mathbb{E}[\int_0^1 \BB(x)^2dx] = \mu_{\mathcal{B}_2} \approx 0.166$. Therefore, the expected asymptotic behavior of the WQT can be approximated as follows,
\begin{align}
     \mathbb{E}\left[D_{wqt}(F_n, G_n)\right] \approx n\pi_1^2 d_{wqt} + \mu_{\mathcal{B}_2}.
\end{align}
\noindent In our simulation and real-world tests, the bias term is removed prior to filtering.

To prove that the WQT is asymptotically symmetric, it suffices to prove that $\int_0^1 \left(F\left(G^{-1}(x)\right)-x\right)^2dx= \int_0^1 \left(G\left(F^{-1}(x)\right)-x\right)^2dx$. This result can be attained by applying the inverse function theorem \cite{key_disks_1994} to $F(G^{-1}(x))^2$ and $G(F^{-1}(x))^2$.

\begin{figure*}
    \centering
    \includegraphics[width=1.5\columnwidth]{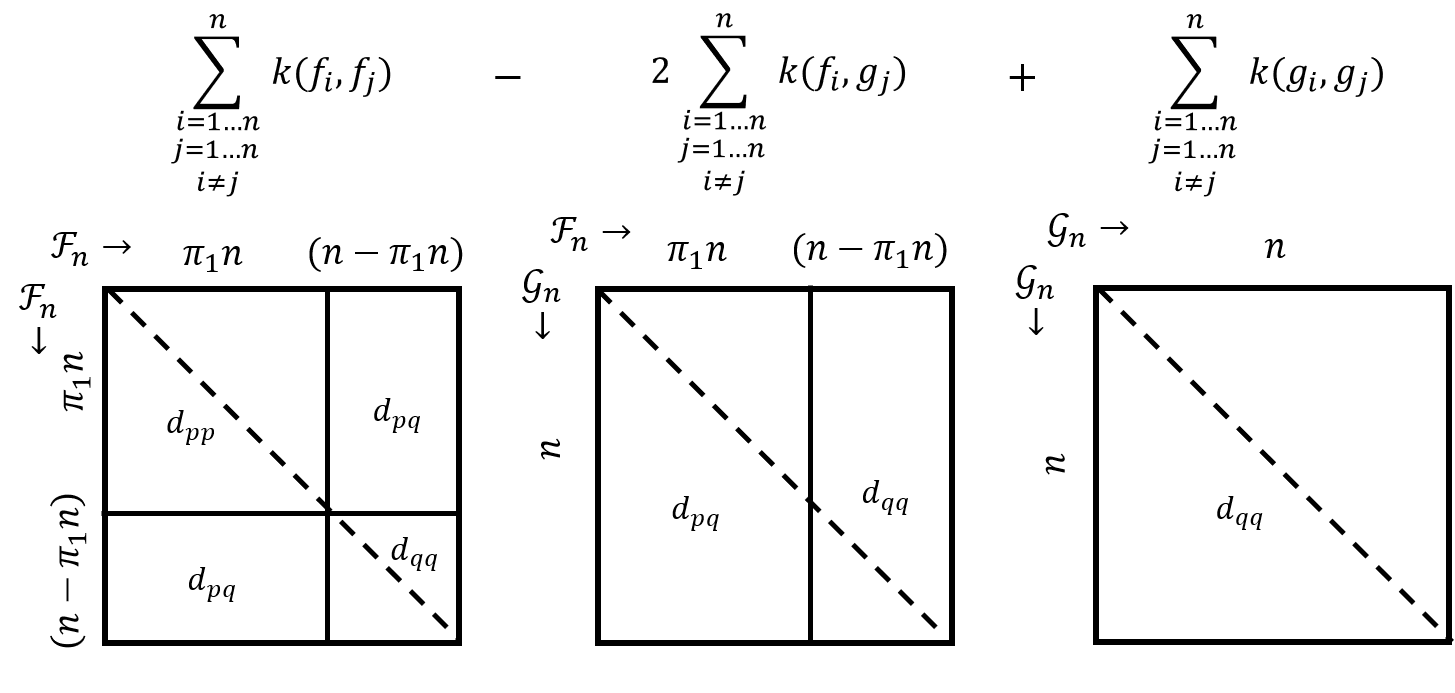}
    \caption{Decomposition of the expected value of the empirical MMD$^2$ between $\mathcal{F}_n$ and $\mathcal{G}_n$ in \eqref{eq:mmd2}. $d_{pp} = \mathbb{E}_{P\times P}[k(p,p')]$, $d_{pq} = \mathbb{E}{P\times Q}[k(p,q')]$, $d_{qq} = \mathbb{E}{Q\times Q}[k(q,q')]$ where $p_i \sim P$, and $q_i \sim Q$}
    \label{fig:mmd}
\end{figure*}

\subsection{Proof of Theorem \hyperref[thm:swqt]{\ThmSWQT}: Sliced Wasserstein Quantile Test} \label{app:SWQT}

\begin{proof}
From the definition of the SWQT, 
\begin{align} \label{eq:swqt}
    \frac{1}{n} &D_{swqt}(\mathcal{F}_n, \mathcal{G}_n) \nonumber \\
    &\triangleq \ints_{\mathcal{S}^{d-1}} \frac{1}{2}\int_0^1 \left(\left( F_n^\theta \circ \left(G_n^\theta\right)^{-1}\right)(x) - x\right)^2dx \,  d\theta
\end{align}
\noindent where $\mathcal{F}_n, \mathcal{G}_n$ denote sets of $n$ IID samples, $F_n^\theta, G_n^\theta$ denotes the one-dimensional CDF attained from projecting the samples of $\mathcal{F}_n, \mathcal{G}_n$ onto the unit vector $\theta$, and $\mathcal{S}^{d-1}$ represents the uniform measure over unit circle in $\mathbb{R}^d$. Once again, given that the distributions of the samples of $\mathcal{F}_n, \mathcal{G}_n$ are $\pi_1 P + (1-\pi_1) Q$ and $Q$ respectively, we apply Thm.~\hyperref[thm:wqt]{\ThmWQT}~to the inner integral,
\begin{align} \label{eq:swqtProof}
    \frac{1}{n} D_{swqt}(&\mathcal{F}_n, \mathcal{G}_n)  \nonumber \\
    &\ra_w \ints_{\mathcal{S}^{d-1}} \frac{\pi_1^2}{2} \int_0^1 \left(\left( P^\theta \circ \left(Q^\theta\right)^{-1}\right)(x) - x\right)^2dx \, d\theta \nonumber \\
    &= \pi_1^2 \ints_{\mathcal{S}^{d-1}}  d_{wqt}(P^\theta,Q^\theta) d\theta \nonumber \\
    &= \pi_1^2 d_{swqt}.
\end{align}

$d_{swqt}$ exists because 1) $P^\theta$, $Q^\theta$ are continuous with respect to $\theta$ because $P$, $Q$ are continuous, and 2) $d_{wqt}(P^\theta, Q^\theta)$ is bounded for all $\theta$ since $P$, $Q$ live on a compact set.  
\end{proof}

By the Portmanteau theorem, $\mathbb{E}[\frac{1}{n} D_{swqt}(\mathcal{F}_n, \mathcal{G}_n)] \ra \pi_1^2 d_{swqt}(\mathcal{F}_n, \mathcal{G}_n)$. In Appx.~\ref{app:WQT} we showed that in the operational case where the window size $n$ is finite, $\mathbb{E}[D_{wqt}(P,Q)] \approx \pi_1^2 d_{wqt}(P,Q) + O(1)$, where the $O(1)$ term arises from the application of Slutsky's theorem. Applying the same reasoning to \eqref{eq:swqtProof}, we have that, 
\begin{align}
    \mathbb{E}[D_{swqt}(&\mathcal{F}_n, \mathcal{G}_n)] \approx \ints_{\mathcal{S}^{d-1}}  n\pi_1^2 d_{wqt}(P^\theta,Q^\theta) + O(1) \, d\theta.
\end{align}

Therefore, $\mathbb{E}[D_{swqt}(\mathcal{F}_n, \mathcal{G}_n)] \approx n \pi_1^{2} d_{swqt}(P,Q) + \int_{\mathcal{S}^{d-1}} O(1) \, d\theta$. We estimate the $O(1)$ term by considering the expected value of the WQT under the null hypothesis that $P=Q$. In this case, $\mathbb{E}[d_{wqt}(P^{\theta},P^{'\theta})]= \mu_{\mathcal{B}_2}$ for all $\theta$, thus it follows that $\int_{\mathcal{S}^{d-1}} O(1) d\theta =\mu_{\mathcal{B}_2}$. This extra factor is treated as a bias that is removed prior to matched filtering.

Therefore, the expected asymptotic behavior of the SWQT can be approximated as follows,
\begin{align}
     \mathbb{E}\left[D_{swqt}(\mathcal{F}_n, \mathcal{G}_n)\right] \approx n\pi_1^2 d_{swqt} + \mu_{\mathcal{B}_2}
\end{align}
Since the WQT is asymptotically symmetric, the SWQT is also asymptotically symmetric.
\subsection{Proof of Theorem \hyperref[thm:ks]{\ThmKS}: Kolmogorov-Smirnov} \label{app:KS}


\begin{proof}
From \eqref{eq:KS}, and the Glivenko-Cantelli theorem (Thm.~\hyperref[thm:gc]{\ThmGC}),
\begin{align}
    D_{KS}(F_n, G_n) = &\sup_x |F_n(x) - G_n(x)| \nonumber \\ 
     \ra_{as} &\sup_x |F(x) - G(x)|.
\end{align}
Then, since $F=\pi_1 P + (1-\pi_1)Q$ and $G=Q$, 
\begin{align}
    D_{KS}(F_n, G_n) \ra_{as}&\pi_1 \sup_x |P(x) - Q(x)| \nonumber \\
    = &\pi_1 d_{KS}(P,Q).
\end{align}
\end{proof}
Therefore, since almost sure convergence implies weak convergence, it follows from the Portmanteau theorem that the expected value of the KS tests converges to,
\begin{align}
     \mathbb{E}\left[D_{KS}(F_n, G_n)\right] \ra \pi_1 d_{KS}(P,Q) 
\end{align}
The symmetric propoerty of the KS tests is easily seen since $|P_n(x)-Q_n(x)|=|Q_n(x)-P_n(x)|$. 


\subsection{Proof of Theorem \hyperref[thm:mmd]{\ThmMMD}: Maximum Mean Discrepancy Squared } \label{app:MMD}


\begin{proof}
Given a positive definite kernel, $k(x,y)$, and following the setup as shown in Fig.~\ref{fig:diagram}, consider two sets of $n$ samples $\mathcal{F}_n = \{f_1, ... f_n\}$ $\mathcal{G}_n = \{g_1, ... , g_n\}$. $\mathcal{F}_n$ will have $\pi_1 n$ samples drawn from $P$ and $\left(n-\pi_1 n\right)$ samples drawn from $Q$ whereas $\mathcal{G}_n$ will have all $n$ samples drawn from $Q$. Since $\pi_1 = \left(1-\frac{t-\tau}{n}\right)$, $\pi_1 n$ is always an integer for integer values of $t$. We denote samples that are drawn from $P$ as $p_i$ and samples drawn from $Q$ as $q_i$. Therefore, 
\begin{align}
 \mathcal{F}_n =& \{p_1, \dots, p_{\pi_1n}, q_{\pi_1 n+1}, \dots, q_n\} \\   
 \mathcal{G}_n =& \{q_1',\dots,q_n'\}.
\end{align}

We decompose the unbiased empirical estimator of the MMD$^2$ distance as stated in \eqref{eq:mmdU} between $\mathcal{F}_n, \mathcal{G}_n$ as visualized in Fig.~\ref{fig:mmd}. 
\begin{align} \label{eq:mmd2}
    &D_{mmd^2}(\mathcal{F}_n, \mathcal{G}_n) \nonumber \\
    &= \frac{1}{n^2-n} \Bigg[ \Bigg(\sum_{\substack{i=1\dots \pi_1 n\\j=1\dots \pi_1 n \\ i\neq j}} k(p_i,p_j) + \sum_{\substack{i=1\dots \pi_1 n\\j=\pi_1 n+1 \dots n}}k(p_i,q_j) \nonumber \\
    &\tab + \sum_{\substack{i=\pi_1 n + 1\dots n\\j=1 \dots \pi_1 n}} k(p_i,q_j)  + \sum_{\substack{i=\pi_1 n+1\dots n\\j=\pi_1 n+1\dots n \\ i\neq j}} k(q_i,q_j)\Bigg) \nonumber \\
    &\tab -  2\left(\sum_{\substack{i=1\dots \pi_1 n\\j=1\dots n \\ i\neq j}} k(p_i,q_j') + \sum_{\substack{i=\pi_1 n+1\dots n\\j=1\dots n \\ i\neq j}} k(q_i,q_j')\right) \nonumber \\
    &\tab + \left(\sum_{\substack{i=1\dots n\\j=1\dots n \\ i\neq j}} k(q_i',q_j')\right)  \Bigg].
\end{align}
We define $\mathbb{E}_{P \times P}[k(p,p')]=d_{pp}$, $\mathbb{E}_{P \times Q}[k(p,q')]= \mathbb{E}_{Q \times P}[k(q,p')]=d_{pq}$, and $\mathbb{E}_{Q \times Q}[k(q',q')]=d_{qq}$. The expectation of the estimator becomes:
\begin{align}
    E\bigg[D_{mmd^2}(\mathcal{F}_n,& \mathcal{G}_n)\bigg] \nonumber \\
    &= \frac{1}{(n^2-n)} \Big(\left[(\pi_1n)^2 -\pi_1n \right]d_{pp} \\ \nonumber 
    & \hspace{3mm}+ \left[2\pi_1(1-\pi_1)n^2 - 2(\pi_1n^2-\pi_1n)\right]d_{pq} \\ \nonumber 
    & \hspace{3mm} + \big[(1-\pi_1)^2n^2-(1-\pi_1)n \nonumber \\
    &\hspace{3mm} - 2((1-\pi_1)n^2-(1-\pi_1)n) + (n^2-n)\big] d_{qq}\Big) \nonumber \\
     &=  \frac{\left(n^2\pi_1^2 - \pi_1 n\right)}{(n^2-n)} \left( d_{pp} + d_{qq} - 2 d_{pq} \right)
\end{align}
\begin{align}
    \lim_{n\rightarrow \infty} E\bigg[D_{mmd^2}(\mathcal{F}_n, \mathcal{G}_n)\bigg]
    &= \pi_1^2 \left( d_{pp} + d_{qq} - 2 d_{pq}\right) \nonumber \\
    &= \pi_1^2 d_{mmd^2}(P,Q).
\end{align}
\end{proof}
The symmetric property of the MMD$^2$ follows from the fact that $k(\cdot, \cdot)$ is symmetric.


\subsection{Maximum Value for Wasserstein Quantile Test}\label{sec:WQTMax}
From the definition of the WQT,
\begin{align}
    D_{wqt}(F_n, G_n) = \frac{n}{2}\ints_0^1 \left(F_n\left(G_n^{-1}(x)\right)\right) -x)^2 dx.
\end{align}
The quantile-quantile map $F(G^{-1}(x))$ is a non-decreasing function where $F(G^{-1}(0))=0$ and $F(G^{-1}(1))=1$. Therefore, the WQT is maximized in two cases: when $F(G^{-1}(x)) \equiv 0$ for all $x\in[0,1)$ or $F(G^{-1}(x))\equiv 1$ for all $x\in(0,1]$. In each case, 
\begin{align}
W_{wqt} = \frac{n}{6}.
\end{align}
This occurs only when the 0-th quantile of $F$ maps to the 100-th quantile of $G$ or vice versa.






\end{appendix}

\end{document}